\documentclass[useAMS,usenatbib]{mn2e}

\usepackage{color}
\usepackage{colortbl}
\usepackage{multirow}
\usepackage{dcolumn}
\usepackage{amsmath}
\usepackage{amssymb}
\usepackage{graphicx}
\usepackage{epsfig}
\usepackage{changebar}
\usepackage{natbib}
\usepackage{multirow}
\usepackage{subfigure}
\usepackage{txfonts}
\usepackage{relsize}

\usepackage[figuresright]{rotating}

\newcolumntype{d}[1]{D{.}{\cdot}{#1}}

\newcolumntype{.}{D{.}{.}{-1}}

\newcommand{\lsun}{L$_\odot$}
\newcommand{\msun}{M$_\odot$}

\newcommand{\vlsr}{V$_{\rm{LSR}}$}

\newcommand{\mum}{$\mu$m}

\newcommand{\kms}{km\,s$^{-1}$}

\newcommand{\HI}{H{\sc i}}
\newcommand{\HII}{H{\sc ii}}

\newcommand{\poi}{Poisson}

\title[Resolving kinematic distance ambiguities towards \HII\
  regions]{The RMS Survey: Resolving kinematic distance ambiguities
  towards a sample of compact \HII\ regions using \HI\ absorption.\thanks{Full versions of Figs.\,3 and 4 are only available in electronic form of the journal.}}
\author[J. S. Urquhart et al.]{
J.\,S.\,Urquhart$^{1,2}$\thanks{E-mail:
jurquhart@mpifr-bonn.mpg.de (MPIfR)}, M.\,G.\,Hoare$^{3}$, S.\,L.\,Lumsden$^{3}$, R.\,D.\,Oudmaijer$^{3}$, T.\,J.\,T.\,Moore$^{4}$, \newauthor J.\,C.\,Mottram$^{5}$, H.\,D.\,B.\,Cooper$^{3}$, M.\,Mottram$^{3}$, H.\,C.\,Rogers$^{3}$ \\
$^{1}$ CSIRO Astronomy and Space Science, P.O. Box\,76, Epping, NSW 1710, Australia \\
$^{2}$ Max-Planck-Institut f\"ur Radioastronomie, Auf dem H\"ugel 69, 53121 Bonn, Germany \\
$^{3}$ School of Physics and Astrophysics, University of Leeds, Leeds, LS2\,9JT, UK \\
$^{4}$ Astrophysics Research Institute, Liverpool John Moores University, Twelve Quays House, Egerton Wharf, Birkenhead, CH41\,1LD, UK\\ 
$^{5}$ School of Physics, University of Exeter, Exeter, EX7\,4QL, UK\\
}

\begin{document}

\date{Accepted ??. Received ??; in original form ??}

\pagerange{\pageref{firstpage}--\pageref{lastpage}} \pubyear{2009}

\maketitle

\label{firstpage}

\begin{abstract}
We present high-resolution \HI\ data obtained using the Australia Telescope Compact Array to resolve the near/far distance ambiguities towards a sample of compact \HII~regions from the Red MSX Source (RMS) survey. The high resolution data are complemented with lower resolution archival \HI\ data extracted from the Southern and VLA Galactic Plane surveys.  We resolve the distance ambiguity for nearly all of the 105 sources where the continuum was strong enough to allow analysis of the \HI\ absorption line structure. This represents another step in the determination of distances to the total RMS sample, which with over 1,000 massive young stellar objects and compact \HII~regions, is the largest and most complete sample of its kind. The full sample will allow the distribution of massive star formation in the Galaxy to be examined.

\end{abstract}

\section{Introduction}

Massive stars ($M_\star>8$\,\msun) are responsible for most of the
energetic phenomena in the Universe.  They deposit large amounts of
radiation, kinetic energy and enriched material into the interstellar
medium during their lives. They may trigger further star formation in
their surrounding environment. These feedback processes play an
important role in regulating star formation within the surrounding
environment, possibly triggering the formation of future generations
of stars, and ultimately driving the evolution of their host galaxy
(\citealt{kennicutt2005}). The specific mechanics of how and where
they form is highly uncertain however. A large scale systematic survey
aimed at identifying and characterising the properties of the massive
young stellar objects (MYSOs) and compact/ultracompact \HII~regions is
required to address these issues. The Red MSX Source (RMS;
\citealt{urquhart2007c}) Survey is designed to return a large,
well-selected sample of young massive stars suited to just this
purpose.

The RMS survey consists of approximately 2000 MYSO candidates spread
throughout the Galaxy ($|b|<5^{\rm o}$) that were identified by
comparing the colours of MSX and 2MASS point sources to those of known
MYSOs (see \citealt{lumsden2002} for details).  In order to
distinguish the MYSOs and ultra-compact (UC) \HII~regions from other
red sources that entered the sample, such as evolved stars and
planetary nebulae (PNe), an ongoing multi-wavelength observational
follow-on programme is being conducted. This includes high resolution
cm continuum observations to identify UC\,\HII~regions and PNe
(\citealt{urquhart_radio_south,urquhart_radio_north}); mid-infrared
imaging to identify genuine point sources, obtain accurate astrometry
and avoid excluding MYSOs located near UC\,\HII~regions
(\citealt{mottram2007}); near-infrared spectroscopy (e.g.,
\citealt{clarke2006}) to distinguish between MYSOs and evolved stars;
and molecular line observations from which we can obtain kinematic
velocities and identify many of the evolved stars that contaminate our
sample (\citealt{urquhart_13co_south,urquhart_13co_north}).

A crucial ingredient in defining the sample and allowing further
 analysis is that the luminosities, and therefore distances, of the
sources are required.  To do this, we use the kinematical distance,
which is the distance derived using a comparatively simple method that
only needs the determination of an object's radial velocity.  This
value can be fitted onto a Galactic rotation curve
\citep[e.g.,][]{brand1993,reid2009} and yields an estimate of that source's
kinematic distance. The radial velocities can be found from Doppler
shifted spectra emitted through the rotational transition of molecular
lines such as from CO, CS or NH$_3$, which can be obtained relatively straightforward
\citep[e.g.,][]{urquhart_13co_south,urquhart_13co_north,urquhart2011b}.

While the source velocity measurement and distance determination in
the outer Galaxy is simple, a problem arises when calculating the
kinematic distances of sources with Galactic radii less than that of
the Sun --- inside the solar circle.  Within the solar circle there
are two possible solutions for every radial velocity, corresponding to
two radial distances. These radial distances are equally spaced on
either side of the tangent point, one on the near side of the object's
orbit and the other on the far side (see Fig.\,1 for schematic
diagram). Only sources actually located at the tangent point avoid
this ambiguity.  This effect is known as the kinematic distance
ambiguity (KDA) and can result in luminosities being calculated with
values that are orders of magnitude in error. However, as young,
massive sources reach the main sequence whilst still embedded in their
natal molecular cloud it is possible to resolve the KDA to these
sources.

There are a number of methods discussed in the literature that can be
applied to do this. For HII
regions which have strong radio continuum emission the basic principle behind solving their KDA is that
interstellar lines such as H$_2$CO (formaldehyde, see e.g.,
\citealt{downes1980,araya2001,araya2002}), or \HI\ absorb (e.g.,
\citealt{kolpak2003}) the continuum free-free emission of the \HII~region. As radial velocities increase to a maximum at the tangent
point along any line of sight, interstellar absorption at velocities
higher than that of the \HII~region implies that the \HII~region
lies at the far distance.  If, on the other hand, the radial velocities
of the absorption lines are smaller than the object's  velocity, then the object
is located at the near distance.  A measure of the object's radial
velocity can be obtained using CO emission from the surrounding cloud
or radio recombination emission lines. This method was used by
\citet{fish2003} and \citet{kolpak2003} to determine the distances to
20 and 49 UC\,\HII~regions respectively.  A similar method using
H110$\alpha$ radio recombination line emission (for the object's
radial velocity) and H$_{2}$CO absorption was used by \citet{Araya},
\citet{watson2003} and \citet{sewilo2004}.

For MYSOs which --- by definition --- do not yet have strong radio
free-free emission (see e.g. Figure 6 in Hoare \& Franco 2007), we can
not use absorption against the target's radio continuum emission.
Methods do exist that use \HI\ self-absorption against any Galactic
background emission instead (e.g.,
\citealt{jackson2002,busfield2006}), but these are much less certain (\citealt{anderson2009a}).

\begin{figure}
\includegraphics[height=0.45\textwidth, trim= 0 0 0 0]{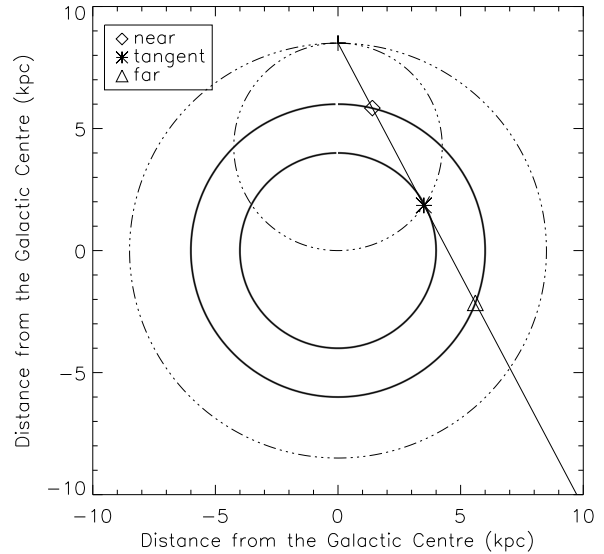}

\caption{\label{fig:gal_tang} Schematic of the KDA. The observer is at (0, 8.5)\,kpc, the line of sight to the target is the solid straight line from this point. Near and far distances are shown from a particular orbit, as is the tangent point at which the radial velocity is just the total velocity. The solar circle is shown by the large dashed circle, and the  locus of the tangent points} by the smaller dashed circle. 
\end{figure}

The RMS survey has identified a sample of $\sim$1300 MYSO candidates
and UC\,\HII~regions in approximately equal numbers and located
throughout the Galaxy. In this paper we focus on resolving the KDA
towards a large sample of UC\,\HII~regions ($\sim$100) located
primarily in the fourth quadrant of the Galaxy using the method
described previously. 

In the next section we briefly describe a set of targeted high
resolution 21\,cm radio observations towards $\sim$80 relatively
bright \HII~regions identified from our previous radio observations
(i.e., \citealt{urquhart_radio_south}). We complement these high
resolution observations with lower resolution archival \HI\ data
extracted from the Southern and VLA Galactic Plane Surveys (SGPS;
\citealt{mcclure2005} and VGPS; \citealt{stil2006}, respectively); an
overview of these surveys is presented in Sect.\,\ref{sect:sgps}. In
Sect.\,\ref{sect:results} we compare the velocity of the \HI\ absorption
seen in the two data sets with molecular line data obtained towards
the RMS sources to resolve the kinematic distance ambiguities towards
these UC\,\HII~regions. We discuss the Galactic distribution of our
sample of young massive stars with respect to the positions of the
spiral arms and Galactic bar in Sect.\,\ref{sect:discussion}.  In
Sect.\,\ref{sect:summary} we summarise our results and present our
conclusions.

\section{\HI\ Data}

\begin{table}


\begin{center}
\caption{Summary of fields observed, synthesised beam parameters and r.m.s. of the restored images.}
\label{tbl:fields}
\begin{minipage}{\linewidth}
\scriptsize
\begin{tabular}{lcccr.}
\hline
\hline
Field Name$^{\rm{a}}$	&RA	& Dec.  & \multicolumn{2}{c}{Beam Parameters} & \multicolumn{1}{c}{Image}  \\

	& J2000	& J2000  & Size & PA & \multicolumn{1}{c}{r.m.s.}  \\
	&(h:m:s) 	& (d:m:s)  & (${\rm{Maj}}\arcsec \times {\rm{Min}}\arcsec$) &(\degr) & \multicolumn{1}{c}{(mJy)}  \\

\hline
G281.0472$-$01.5432	&	09:59:16.51	&	$-$56:54:43.2	&	$12.0\times5.7$	&	$-$74	&	0.8	\\
G281.5576$-$02.4775	&	09:58:02.85	&	$-$57:57:48.9	&	$12.1\times5.6$	&	$-$78	&	0.6	\\
G281.8449$-$01.6094	&	10:03:40.96	&	$-$57:26:39.8	&	$11.9\times5.7$	&	$-$78	&	0.5	\\
G283.2273$-$00.9353	&	10:15:00.07	&	$-$57:41:38.7	&	$11.9\times5.7$	&	$-$78	&	0.4	\\
G305.1967+00.0335	&	13:11:14.61	&	$-$62:45:04.3	&	$7.6\times6.6$	&	3	&	2.3	\\
G305.2535+00.2412$^\star$	&	13:11:35.80	&	$-$62:32:22.9	&	$7.6\times6.7$	&	4	&	1.9	\\
G307.5606$-$00.5871	&	13:32:31.15	&	$-$63:05:21.1	&	$7.6\times6.6$	&	7	&	0.3	\\
G307.6213$-$00.2622	&	13:32:35.49	&	$-$62:45:31.3	&	$7.7\times6.6$	&	1	&	0.6	\\
G308.0023+02.0190	&	13:32:42.00	&	$-$60:26:45.2	&	$8.1\times6.4$	&	1	&	0.2	\\
G311.1794$-$00.0720	&	14:02:08.44	&	$-$61:48:23.4	&	$7.8\times6.5$	&	2	&	0.3	\\
G311.4255+00.5964	&	14:02:36.38	&	$-$61:05:46.6	&	$7.8\times6.6$	&	3	&	0.6	\\
G312.3070+00.6613	&	14:09:24.79	&	$-$60:46:59.5	&	$7.9\times6.5$	&	2	&	0.5	\\
G312.5472$-$00.2801	&	14:13:41.71	&	$-$61:36:24.4	&	$7.8\times6.5$	&	1	&	0.3	\\
G314.2161+00.2546$^\star$	&	14:25:14.04	&	$-$60:32:44.8	&	$7.4\times6.8$	&	$-$31	&	0.3	\\
G314.2204+00.2726	&	14:25:12.88	&	$-$60:31:38.6	&	$7.4\times6.9$	&	$-$33	&	0.4	\\
G316.1386$-$00.5009	&	14:42:01.58	&	$-$60:30:20.1	&	$7.5\times6.9$	&	$-$27	&	0.4	\\
G318.7251$-$00.2241	&	14:59:30.09	&	$-$59:06:41.7	&	$7.4\times7.0$	&	$-$23	&	0.3	\\
G318.9148$-$00.1647	&	15:00:34.94	&	$-$58:58:10.2	&	$7.7\times7.2$	&	$-$24	&	0.9	\\
G325.5159+00.4147$^\star$	&	15:39:11.21	&	$-$54:55:36.8	&	$10.1\times6.3$	&	88	&	0.3	\\
G326.4719$-$00.3777	&	15:47:49.80	&	$-$54:58:34.3	&	$10.2\times6.2$	&	87	&	0.4	\\
G326.7249+00.6159	&	15:44:59.44	&	$-$54:02:13.9	&	$8.2\times6.9$	&	$-$24	&	1.2	\\
G327.4014+00.4454	&	15:49:19.03	&	$-$53:45:12.9	&	$8.1\times6.9$	&	$-$31	&	0.3	\\
G327.8483+00.0175	&	15:53:29.47	&	$-$53:48:18.0	&	$8.1\times6.8$	&	$-$21	&	0.4	\\
G328.3067+00.4308	&	15:54:06.23	&	$-$53:11:40.2	&	$8.1\times7.1$	&	$-$22	&	0.5	\\
G329.4761+00.8414	&	15:58:16.53	&	$-$52:07:43.3	&	$9.9\times6.3$	&	57	&	0.4	\\
G329.5982+00.0560	&	16:02:14.59	&	$-$52:38:37.6	&	$9.8\times6.4$	&	60	&	0.9	\\
G330.2845+00.4933	&	16:03:43.46	&	$-$51:51:44.2	&	$9.7\times6.5$	&	61	&	0.2	\\
G330.2935$-$00.3946	&	16:07:38.06	&	$-$52:31:03.7	&	$9.2\times6.6$	&	72	&	0.4	\\
G330.9544$-$00.1817	&	16:09:52.77	&	$-$51:54:52.2	&	$10.3\times6.4$	&	55	&	0.5	\\
G331.1465+00.1343	&	16:09:24.55	&	$-$51:33:06.8	&	$9.4\times6.6$	&	67	&	0.7	\\
G331.3546+01.0638	&	16:06:24.16	&	$-$50:43:27.4	&	$9.6\times6.6$	&	64	&	0.5	\\
G331.4181$-$00.3546	&	16:12:50.25	&	$-$51:43:29.9	&	$9.2\times6.2$	&	1	&	1.0	\\
G331.4904$-$00.1173$^\star$	&	16:12:07.84	&	$-$51:30:09.0	&	$8.8\times6.8$	&	$-$10	&	0.6	\\
G332.2944$-$00.0962	&	16:15:45.86	&	$-$50:56:02.3	&	$8.3\times7.0$	&	$-$19	&	0.7	\\
G332.5438$-$00.1277	&	16:17:02.47	&	$-$50:47:00.9	&	$8.5\times6.8$	&	$-$17	&	0.6	\\
G332.8256$-$00.5498	&	16:20:11.18	&	$-$50:53:17.5	&	$9.3\times6.3$	&	$-$1	&	1.0	\\
G333.0162+00.7615	&	16:15:18.64	&	$-$49:48:55.0	&	$8.6\times6.8$	&	$-$13	&	0.5	\\
G333.3072$-$00.3666	&	16:21:31.63	&	$-$50:25:08.0	&	$10.3\times5.8$	&	9	&	2.4	\\
G333.6788$-$00.4344	&	16:23:28.22	&	$-$50:12:12.2	&	$9.7\times6.4$	&	1	&	1.2	\\
G335.1972$-$00.3884	&	16:29:47.59	&	$-$49:04:51.2	&	$9.5\times6.4$	&	$-$10	&	0.3	\\
G335.5783$-$00.2075	&	16:30:35.28	&	$-$48:40:47.2	&	$8.3\times7.3$	&	11	&	0.3	\\
G336.8877+00.0483$^\star$	&	16:34:48.72	&	$-$47:32:49.5	&	$9.5\times7.1$	&	68	&	2.3	\\
G336.9920$-$00.0244	&	16:35:32.83	&	$-$47:31:09.8	&	$8.4\times7.5$	&	56	&	1.4	\\
G337.0047+00.3226	&	16:34:05.25	&	$-$47:16:30.7	&	$11.0\times5.8$	&	$-$3	&	0.5	\\
G337.4050$-$00.4071	&	16:38:52.03	&	$-$47:28:11.2	&	$10.1\times6.1$	&	$-$10	&	0.9	\\
G337.6651$-$00.1750	&	16:38:52.22	&	$-$47:07:16.3	&	$10.3\times6.3$	&	$-$9	&	1.3	\\
G337.7091+00.0932	&	16:37:52.29	&	$-$46:54:33.1	&	$10.1\times6.2$	&	$-$12	&	0.7	\\
G338.2900$-$00.3729	&	16:42:09.98	&	$-$46:47:04.2	&	$10.2\times6.2$	&	$-$9	&	1.1	\\
G338.3340+00.1315	&	16:40:07.96	&	$-$46:25:04.0	&	$10.2\times6.3$	&	$-$11	&	1.2	\\
G338.6811$-$00.0844	&	16:42:24.14	&	$-$46:18:00.7	&	$10.3\times6.2$	&	$-$10	&	1.0	\\
G338.9173+00.3824	&	16:41:16.65	&	$-$45:48:52.9	&	$10.3\times6.2$	&	$-$8	&	0.3	\\
G339.1052+00.1490	&	16:42:59.81	&	$-$45:49:37.9	&	$10.5\times6.2$	&	$-$11	&	0.4	\\
G339.9797$-$00.5391	&	16:49:14.90	&	$-$45:36:34.2	&	$10.4\times6.2$	&	$-$9	&	0.3	\\
G340.2480$-$00.3725	&	16:49:30.14	&	$-$45:17:48.4	&	$10.5\times6.2$	&	$-$9	&	0.2	\\
G340.2490$-$00.0460	&	16:48:05.25	&	$-$45:05:09.6	&	$10.5\times6.2$	&	$-$8	&	0.4	\\
G344.2207$-$00.5953	&	17:04:13.32	&	$-$42:19:57.3	&	$10.7\times6.5$	&	$-$2	&	0.3	\\
G344.4257+00.0451	&	17:02:09.65	&	$-$41:46:46.2	&	$11.0\times6.5$	&	$-$2	&	0.3	\\
G345.0034$-$00.2240	&	17:05:11.16	&	$-$41:29:06.0	&	$10.6\times6.4$	&	$-$2	&	0.2	\\
G345.4881+00.3148	&	17:04:28.17	&	$-$40:46:22.4	&	$10.9\times6.5$	&	$-$2	&	0.8	\\
G345.5472$-$00.0801	&	17:06:19.34	&	$-$40:57:52.9	&	$10.9\times6.3$	&	$-$2	&	0.8	\\
G345.6495+00.0084	&	17:06:16.48	&	$-$40:49:46.9	&	$11.0\times6.4$	&	$-$2	&	0.8	\\
G346.5235+00.0839	&	17:08:42.83	&	$-$40:05:06.3	&	$10.9\times6.4$	&	$-$2	&	0.4	\\
G347.2326+01.2633	&	17:06:01.89	&	$-$38:48:36.3	&	$10.9\times6.6$	&	20	&	0.2	\\
G347.5998+00.2442	&	17:11:21.91	&	$-$39:07:27.1	&	$10.4\times6.9$	&	23	&	0.6	\\
G348.5312$-$00.9714	&	17:19:15.28	&	$-$39:04:31.0	&	$10.9\times6.6$	&	24	&	1.0	\\
G348.6972$-$01.0263	&	17:19:58.55	&	$-$38:58:14.5	&	$11.0\times7.3$	&	41	&	0.8	\\
G348.8922$-$00.1787	&	17:16:59.99	&	$-$38:19:24.6	&	$10.8\times6.6$	&	22	&	0.4	\\
G349.1055+00.1121	&	17:16:25.39	&	$-$37:58:51.9	&	$10.7\times6.8$	&	23	&	0.5	\\
G349.7215+00.1203$^\star$	&	17:18:11.39	&	$-$37:28:25.3	&	$11.1\times6.5$	&	21	&	1.5	\\
\hline\\
\end{tabular}\\
$^{\rm{a}}$ We identify sources towards which no radio emission is detected by appending a $\star$ to the field name.
\end{minipage}
\end{center}
\end{table}

\subsection{Compact Array Observations}

\subsubsection{Description of set up and procedures}

Observations were made using the Australia Telescope Compact Array (ATCA) between the 24th and 28th of January 2008 (Project code C1772; \citealt{lumsden2007}). The ATCA is located at the Paul Wild Observatory, Narrabri, New South Wales, Australia and consists of $6\times22$\,m antennas, 5 of which lie on a 3\,km east-west railway track with the sixth antenna located 3\,km farther west. This allows the antennas to be positioned in several configurations with baselines ranging in length from 30\,metres up to 6\,km. 

A 6-km array configuration was used to achieve an effective spatial resolution for our sources of about 10\arcsec\ at 21\,cm. The correlator was set up to make simultaneous continuum and spectral line observations at 21\,cm; the continuum observations used a bandpass of 128\,MHz centred at a frequency of 1384\,MHz, while the spectral line observations used 8\,MHz of bandwidth with 512 spectral channels centred at 1422\,MHz to detect neutral hydrogen (\HI) absorption. With an 8\,MHz bandwidth and 512 channels the spectral line observations provides a velocity range of approximately 1600\,\kms\ with a channel resolution of $\sim$3.3\,\kms.

In total 69 fields were observed which include 85 compact \HII~regions
identified by the RMS survey. We have chosen targets in the range
27-200\,mJy to observe from the list of Urquhart et al. (2007).
Fields were grouped by position into small blocks of between 8-10
sources; these were observed in snapshot mode, which consisted of 5-6
cuts of 10 minutes ($\sim$60 minutes total on-source integration)
separated over a 10-12 hour period to provide good
\emph{uv}-coverage. To correct for fluctuations in the phase and
amplitude of these data, caused by atmospheric and instrumental
effects, each block was sandwiched between two short observations of a
nearby phase calibrator (typically 2-3 minutes depending on the flux
density of the calibrator). To allow the absolute calibration of the
flux density and bandpass the primary flux calibrator 1934$-$638 was
observed once each day. The theoretical continuum and spectral line
sensitivity of these observations is 0.1\,mJy beam$^{-1}$ and 8\,mJy
beam$^{-1}$ channel$^{-1}$, however, due to the limited
\emph{uv}-coverage the dynamic range limits the final sensitivity to a
few times this.

The field names and positions are presented in Table\,1 along with the parameters of the restoring beam and 1$\sigma_{\rm{r.m.s}}$ noise measurements obtained from emission free regions in the final continuum maps of each field.

\subsubsection{Data reduction}

The calibration and reduction of these data were performed using the MIRIAD reduction package \citep{sault1995} following standard ATCA procedures. Maps were made of the continuum emission out to the the FWHM of the primary beam (i.e., $\sim$30$^{\prime}$ at 21\,cm). The image pixel size was chosen to provide $\sim$3 pixels across the synthesised beam (in this case 2\arcsec). A robust weighting of 0.5 was used in the  deconvolution as it produces images with the same sensitivity as natural weighting, but with a much improved beam-shape and lower sidelobe contamination. These maps were CLEANed using up to a few thousand cleaning components, or until the residuals were less than three times the theoretical noise value.

\begin{figure*}
\begin{center}
\includegraphics[width=0.33\textwidth, trim= 30 0 0 0]{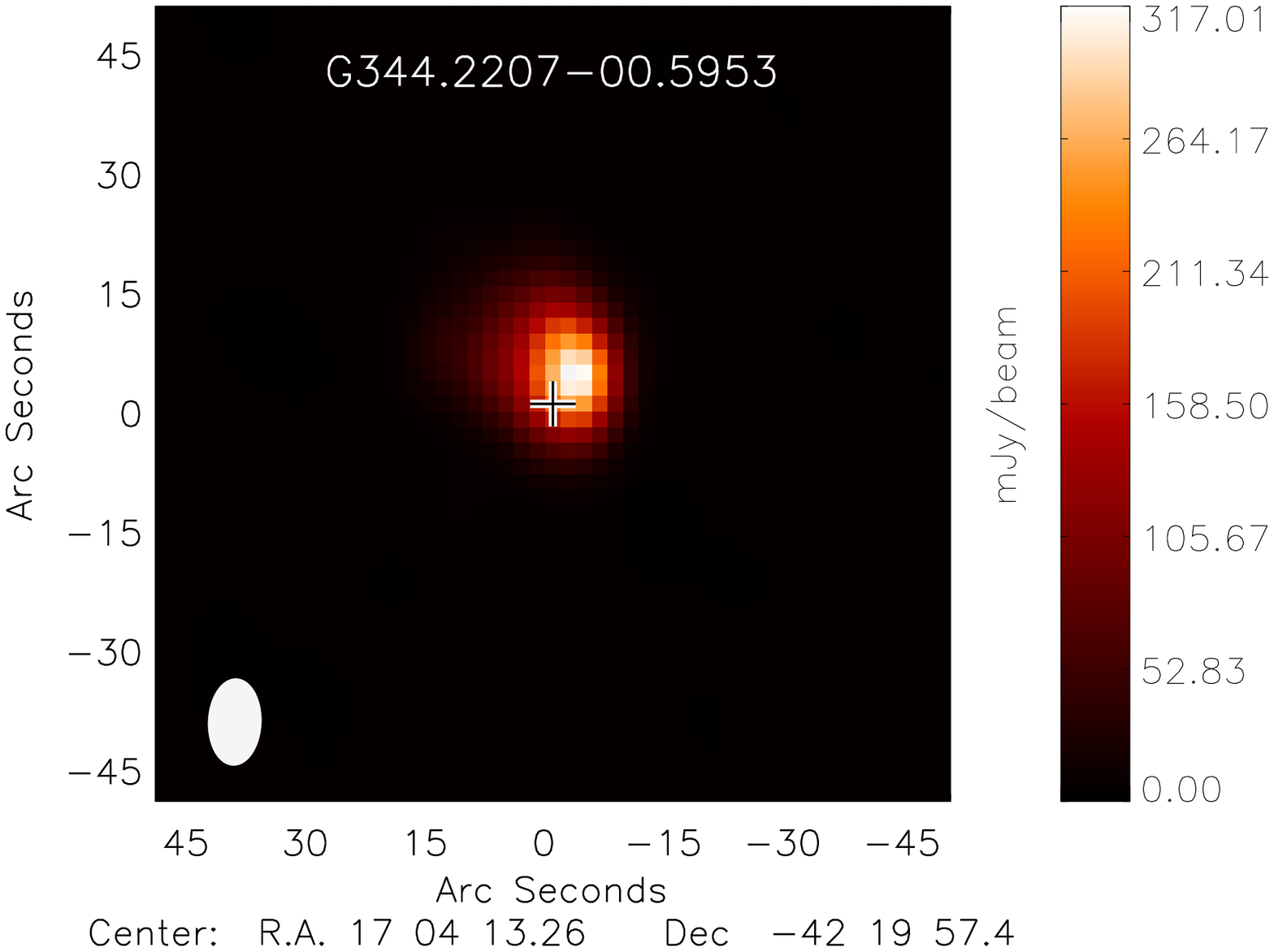}
\includegraphics[width=0.33\textwidth, trim= 30 0 0 0]{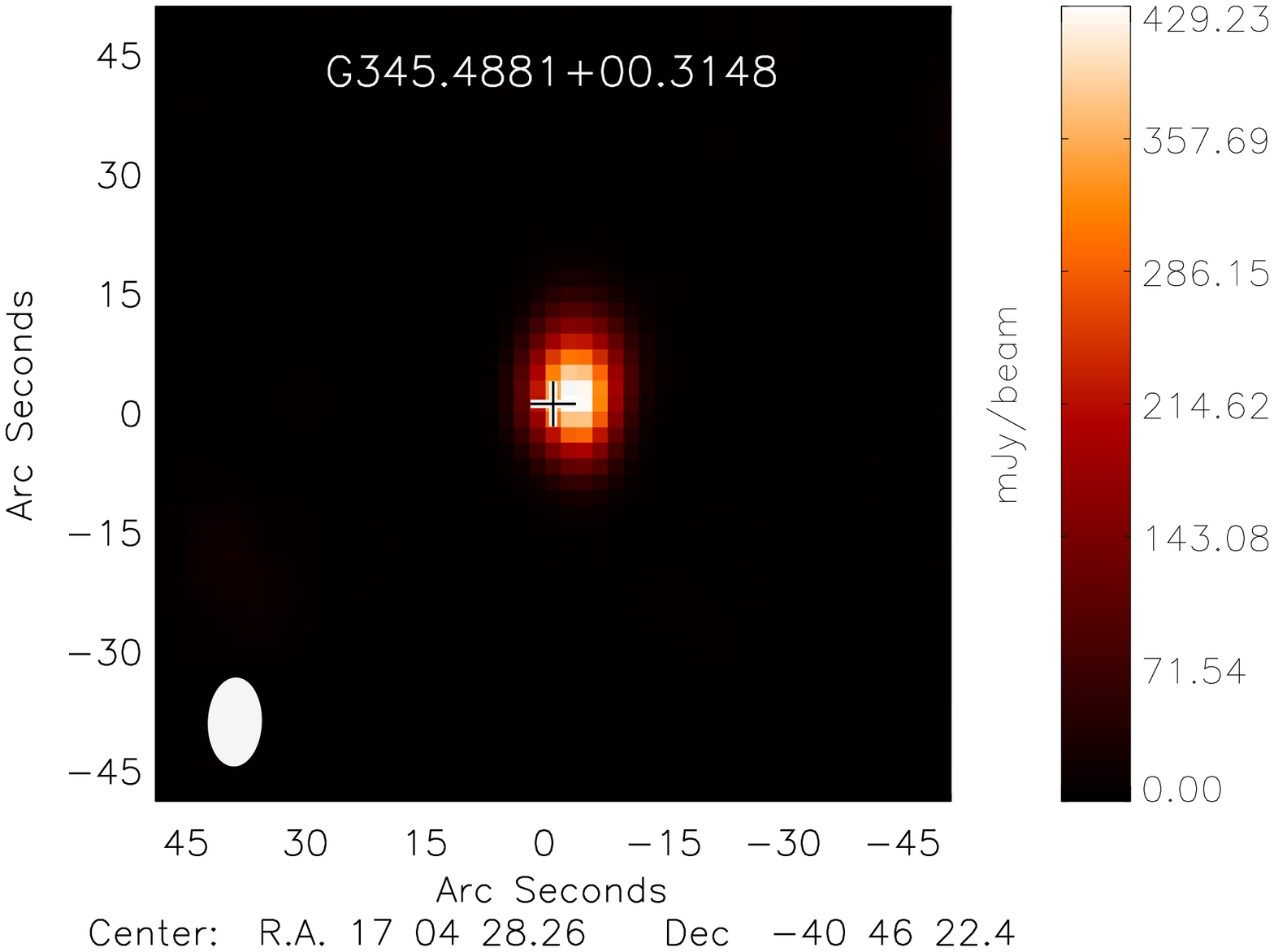}
\includegraphics[width=0.33\textwidth, trim= 30 0 0 0]{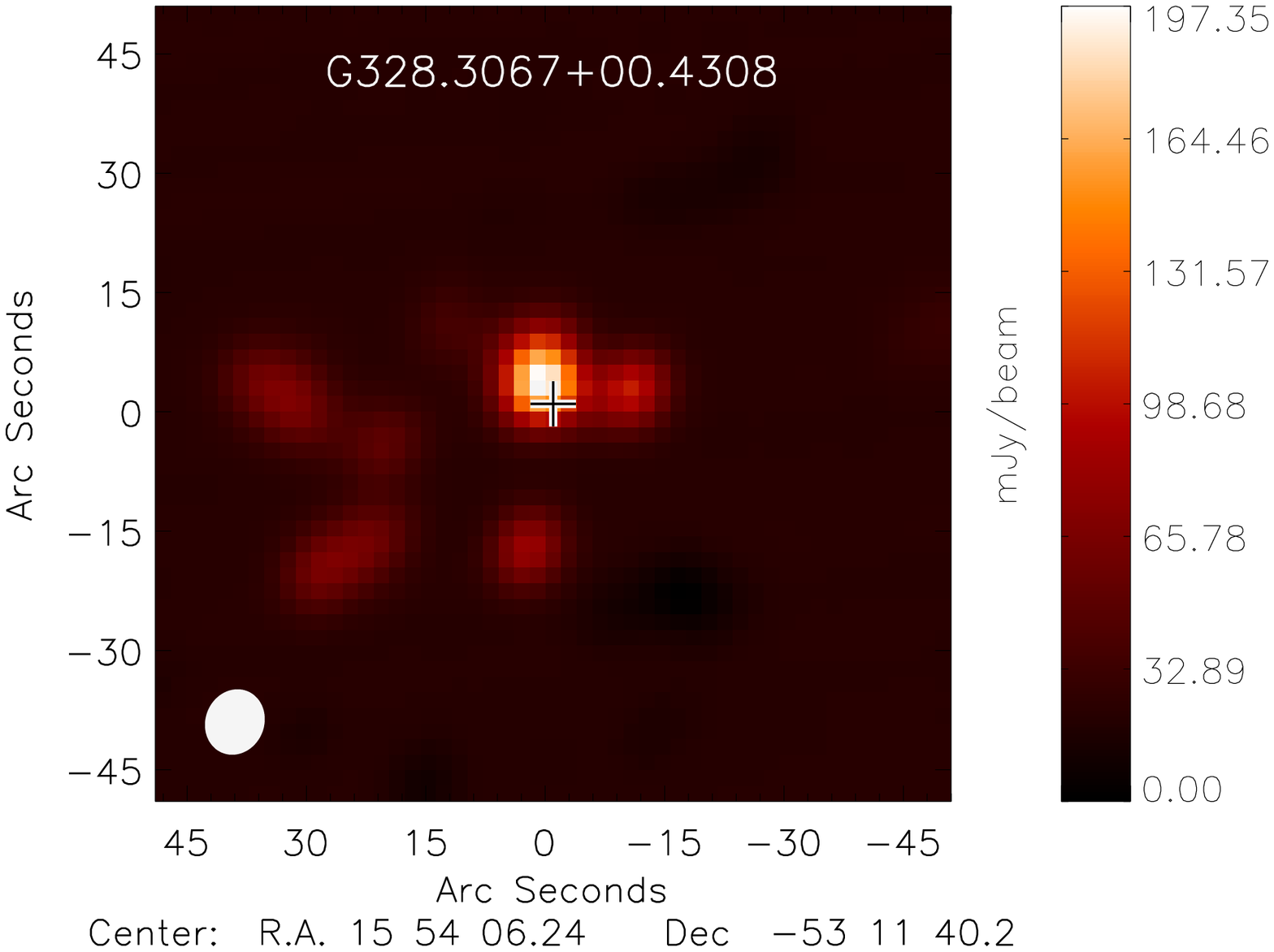}
\includegraphics[width=0.33\textwidth, trim= 30 0 0 0]{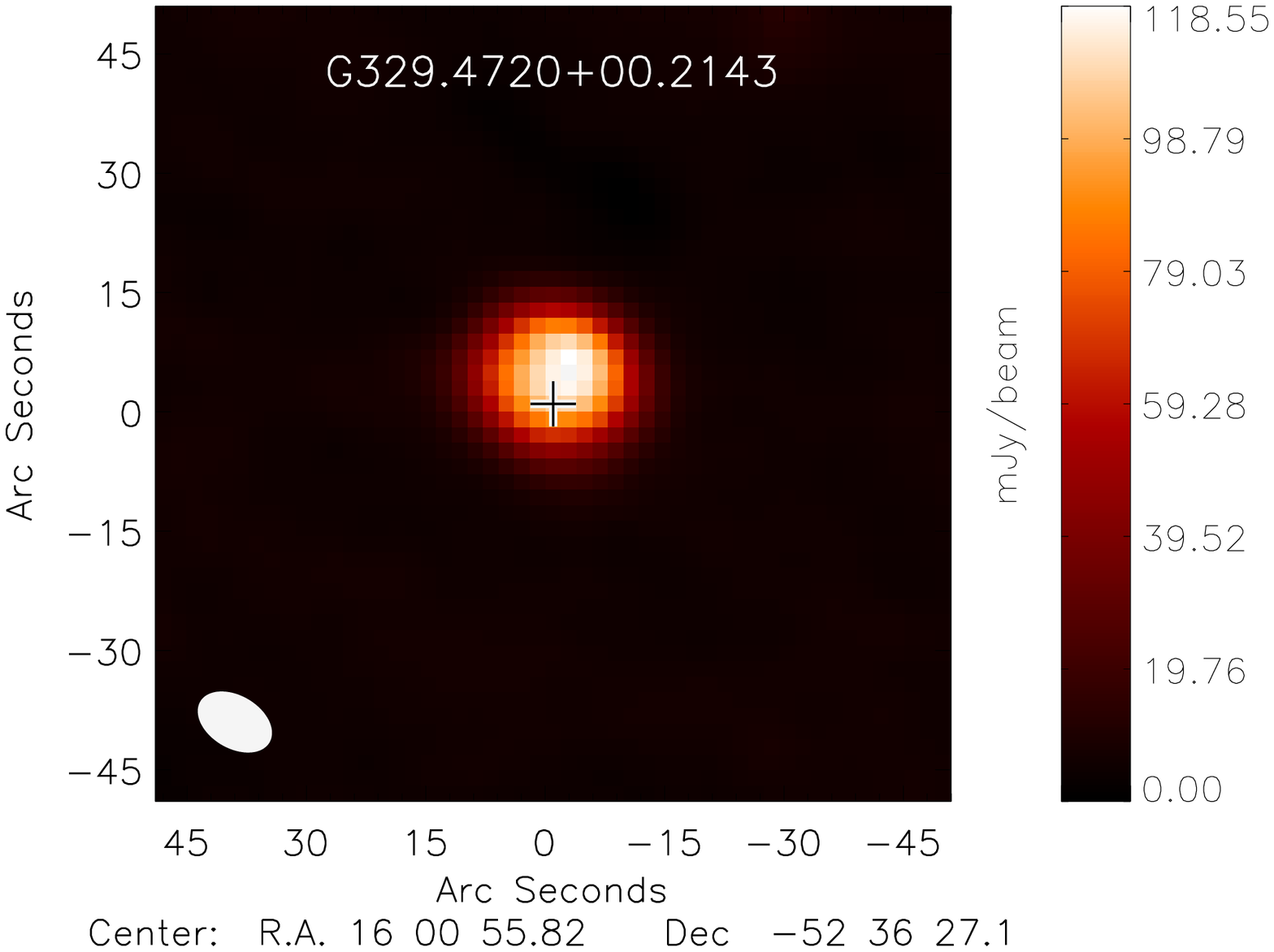}
\includegraphics[width=0.33\textwidth, trim= 30 0 0 0]{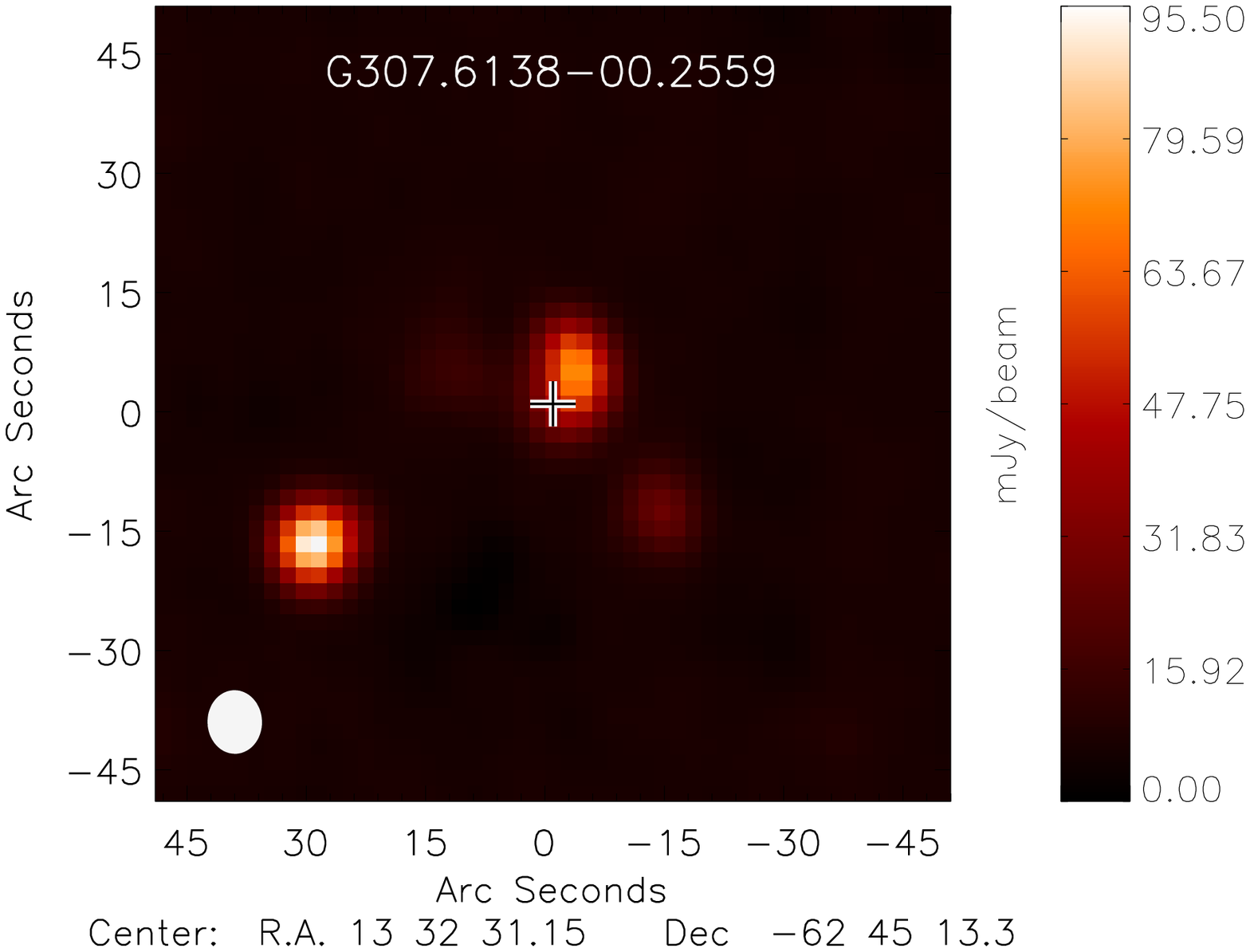}
\includegraphics[width=0.33\textwidth, trim= 30 0 0 0]{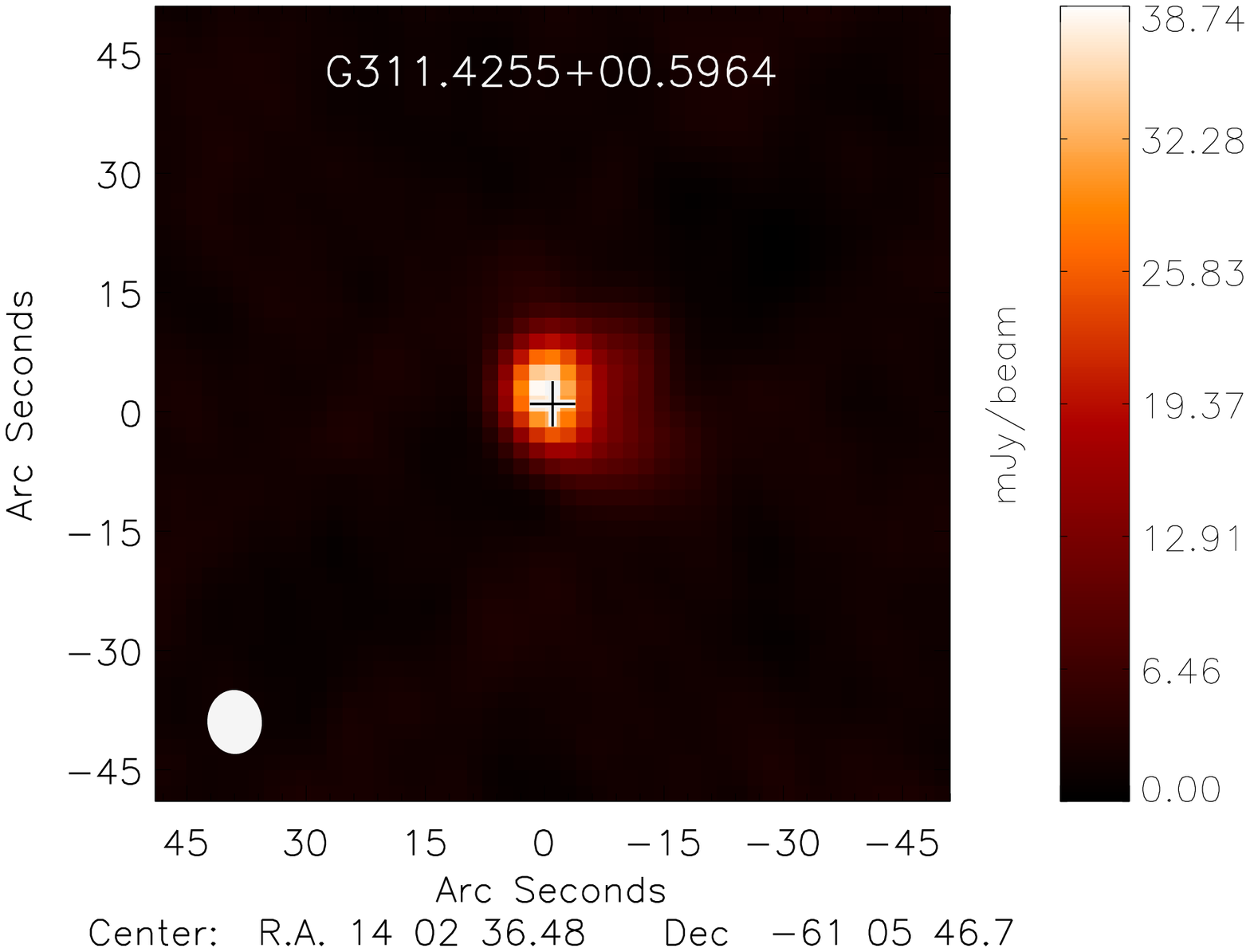}

\caption{\label{fig:example_images} Radio 21\,cm continuum emission maps of a sample of targeted \HII~regions. The size of the synthesised beam is indicated by the white ellipse shown in the lower left corner of each map and the position of the MSX source is indicated by a cross.} 

\end{center}
\end{figure*}

These images were then examined for compact, high surface brightness sources using a nominal 4$\sigma_{\rm{r.m.s}}$ detection threshold, where $\sigma_{\rm{r.m.s}}$ refers to the image r.m.s. noise level. No emission was detected within six fields, with multiple sources detected in 12 fields and a single source being detected in the remaining 52 fields. In Table\,\ref{tbl:fields} we indicate the fields with no detected radio emission by appending a $\star$ to the field name. In Fig.\,\ref{fig:example_images} we present emission maps for a sample of the radio detections and tabulate the source parameters in Table\,\ref{tbl:radio_detectionsfields}.

\setlength{\tabcolsep}{3pt}
\begin{table}


\begin{center}
\caption{Radio source parameters.}
\label{tbl:radio_detectionsfields}
\begin{minipage}{\linewidth}
\scriptsize
\begin{tabular}{lcc..cr}
\hline
\hline
Field Name	&\multicolumn{2}{c}{Position (J2000)}  & \multicolumn{2}{c}{Continuum Flux} & \multicolumn{2}{c}{Source Size} \\

	& RA	& Dec.  & \multicolumn{1}{c}{Peak} & \multicolumn{1}{c}{Int.} & Maj$\times {\rm{Min}}$ & \multicolumn{1}{c}{PA}    \\
	&\multicolumn{1}{c}{(h:m:s)} 	& \multicolumn{1}{c}{(d:m:s)}  & \multicolumn{1}{c}{(mJy)} &\multicolumn{1}{c}{(mJy)} &\multicolumn{1}{c}{(\arcsec)} &\multicolumn{1}{c}{(\degr)}   \\

\hline

G281.0472$-$01.5432	&	9:59:16.727	&	$-$56:54:39.96	&	52.2	&	284.5	&	$20.2\times13.5$	&	$-$8	\\
G281.5576$-$02.4775	&	9:58:02.979	&	$-$57:57:45.41	&	280.6	&	526.3	&	$8.0\times6.2$	&	$-$29	\\
G281.8449$-$01.6094	&	10:03:40.994	&	$-$57:26:39.52	&	54.4	&	112.3	&	$9.2\times5.2$	&	23	\\
G283.2273$-$00.9353	&	10:14:59.782	&	$-$57:41:37.77	&	13.7	&	15.5	&	$3.8\times2.2$	&	$-$67	\\
G305.1967+00.0335	&	13:11:14.440	&	$-$62:45:01.82	&	377.3	&	2127.0	&	$17.0\times13.7$	&	52	\\
G305.2694$-$00.0072	&	13:11:54.915	&	$-$62:47:08.20	&	241.6	&	1392.0	&	$19.6\times12.1$	&	29	\\
G305.3500+00.2240	&	13:12:26.343	&	$-$62:32:59.84	&	102.7	&	415.5	&	$15.4\times9.8$	&	36	\\
G307.5606$-$00.5871	&	13:32:31.011	&	$-$63:05:20.66	&	262.3	&	613.6	&	$8.5\times7.8$	&	$-$65	\\
G307.6138$-$00.2559	&	13:32:30.772	&	$-$62:45:09.65	&	72.3	&	137.6	&	$8.5\times5.2$	&	$-$1	\\
G307.6213$-$00.2622	&	13:32:35.591	&	$-$62:45:31.21	&	97.2	&	135.6	&	$5.1\times3.6$	&	$-$73	\\
G308.0023+02.0190	&	13:32:42.200	&	$-$60:26:45.71	&	14.6	&	16.1	&	$2.5\times2.1$	&	52	\\
G311.1794$-$00.0720	&	14:02:08.212	&	$-$61:48:24.77	&	7.0	&	7.6	&	$2.7\times0.7$	&	$-$69	\\
G311.4255+00.5964	&	14:02:36.408	&	$-$61:05:45.40	&	35.8	&	111.2	&	$11.3\times9.5$	&	32	\\
G312.3070+00.6613	&	14:09:25.002	&	$-$60:47:01.80	&	6.2	&	8.1	&	$5.1\times2.8$	&	23	\\
G312.5472$-$00.2801	&	14:13:41.996	&	$-$61:36:26.44	&	15.3	&	15.9	&	$2.0\times0.7$	&	$-$46	\\
G314.2204+00.2726	&	14:25:12.690	&	$-$60:31:38.72	&	5.0	&	5.8	&	$3.5\times1.9$	&	$-$78	\\
G316.1386$-$00.5009	&	14:42:01.687	&	$-$60:30:22.99	&	129.2	&	966.0	&	$20.1\times16.5$	&	16	\\
G318.7251$-$00.2241	&	14:59:29.802	&	$-$59:06:36.70	&	89.0	&	125.3	&	$4.8\times4.3$	&	$-$35	\\
G318.7748$-$00.1513	&	14:59:34.575	&	$-$59:01:23.94	&	57.7	&	245.4	&	$13.9\times12.0$	&	85	\\
G318.9148$-$00.1647	&	15:00:34.705	&	$-$58:58:09.20	&	431.3	&	970.9	&	$10.5\times6.3$	&	$-$28	\\
G326.4719$-$00.3777	&	15:47:49.858	&	$-$54:58:32.17	&	118.4	&	152.7	&	$4.5\times3.1$	&	14	\\
G326.7249+00.6159	&	15:44:59.314	&	$-$54:02:14.83	&	364.1	&	1443.0	&	$13.2\times12.5$	&	$-$84	\\
G327.4014+00.4454	&	15:49:19.332	&	$-$53:45:13.19	&	23.7	&	26.1	&	$2.5\times2.3$	&	70	\\
G327.8483+00.0175	&	15:53:29.299	&	$-$53:48:16.99	&	22.4	&	24.5	&	$2.4\times2.1$	&	77	\\
G328.3067+00.4308	&	15:54:06.272	&	$-$53:11:37.15	&	168.4	&	404.7	&	$12.4\times5.2$	&	76	\\
G329.4720+00.2143	&	16:00:55.748	&	$-$52:36:23.57	&	132.2	&	479.8	&	$13.4\times11.8$	&	$-$59	\\
G329.4761+00.8414	&	15:58:16.606	&	$-$52:07:37.40	&	21.1	&	59.9	&	$11.8\times9.0$	&	$-$4	\\
G329.5982+00.0560	&	16:02:14.422	&	$-$52:38:32.62	&	38.2	&	276.0	&	$21.2\times17.8$	&	$-$16	\\
G330.2845+00.4933	&	16:03:43.298	&	$-$51:51:45.77	&	29.2	&	31.2	&	$2.5\times1.6$	&	73	\\
G330.2935$-$00.3946	&	16:07:37.810	&	$-$52:31:02.46	&	87.2	&	319.1	&	$14.7\times10.4$	&	9	\\
G330.9544$-$00.1817	&	16:09:52.668	&	$-$51:54:52.62	&	107.9	&	225.2	&	$15.3\times1.7$	&	23	\\
G331.1465+00.1343	&	16:09:24.233	&	$-$51:33:07.48	&	44.3	&	97.6	&	$9.3\times7.5$	&	$-$7	\\
G331.3546+01.0638	&	16:06:26.126	&	$-$50:43:17.82	&	54.4	&	813.0	&	$42.2\times20.9$	&	70	\\
G331.3865$-$00.3598	&	16:12:42.557	&	$-$51:45:00.76	&	25.1	&	91.4	&	$15.6\times9.3$	&	31	\\
G331.4181$-$00.3546	&	16:12:50.260	&	$-$51:43:29.67	&	87.6	&	109.2	&	$4.1\times2.6$	&	81	\\
G332.2944$-$00.0962	&	16:15:45.886	&	$-$50:56:03.42	&	57.4	&	123.4	&	$10.5\times5.5$	&	77	\\
G332.5438$-$00.1277	&	16:17:02.327	&	$-$50:47:02.51	&	39.0	&	139.8	&	$12.5\times11.9$	&	48	\\
G332.8256$-$00.5498	&	16:20:11.037	&	$-$50:53:19.47	&	111.3	&	578.0	&	$24.4\times9.1$	&	36	\\
G333.0162+00.7615	&	16:15:18.651	&	$-$49:48:52.71	&	33.0	&	47.9	&	$5.9\times4.0$	&	$-$89	\\
G333.1306$-$00.4275	&	16:21:00.071	&	$-$50:35:09.24	&	160.4	&	560.0	&	$13.6\times9.5$	&	$-$73	\\
G333.2880$-$00.3907	&	16:21:31.619	&	$-$50:26:59.89	&	307.6	&	938.9	&	$12.0\times8.9$	&	$-$79	\\
G333.3072$-$00.3666	&	16:21:31.588	&	$-$50:25:05.93	&	221.2	&	1342.0	&	$22.8\times12.8$	&	37	\\
G333.6032$-$00.2184	&	16:22:09.555	&	$-$50:06:01.41	&	235.3	&	3061.0	&	$35.4\times21.0$	&	$-$23	\\
G333.6788$-$00.4344	&	16:23:28.338	&	$-$50:12:14.00	&	33.8	&	37.7	&	$3.6\times1.5$	&	35	\\
G335.1972$-$00.3884	&	16:29:47.340	&	$-$49:04:47.88	&	20.4	&	53.5	&	$10.9\times8.6$	&	37	\\
G335.5783$-$00.2075	&	16:30:34.901	&	$-$48:40:46.79	&	30.8	&	71.2	&	$12.1\times6.2$	&	4	\\
G336.9920$-$00.0244	&	16:35:32.289	&	$-$47:31:13.37	&	79.3	&	122.6	&	$6.1\times5.6$	&	$-$12	\\
G337.0047+00.3226	&	16:34:04.700	&	$-$47:16:29.46	&	125.4	&	168.4	&	$4.7\times3.4$	&	72	\\
G337.1218$-$00.1748	&	16:36:42.498	&	$-$47:31:30.47	&	195.3	&	1392.0	&	$22.4\times17.2$	&	60	\\
G337.4050$-$00.4071	&	16:38:50.465	&	$-$47:28.03.59	&	22.1	&	25.2	&	$0.0\times0.0$	&	0	\\
G337.6651$-$00.1750	&	16:38:52.383	&	$-$47:07:16.96	&	202.1	&	366.1	&	$7.4\times6.3$	&	$-$78	\\
G337.7051$-$00.0575	&	16:38:30.878	&	$-$47:00:46.93	&	46.8	&	131.0	&	$11.6\times9.5$	&	37	\\
G337.7091+00.0932	&	16:37:51.954	&	$-$46:54:33.47	&	112.8	&	191.1	&	$6.5\times6.2$	&	62	\\
G338.2900$-$00.3729	&	16:42:09.314	&	$-$46:47:02.87	&	29.5	&	43.2	&	$6.0\times3.9$	&	57	\\
G338.3340+00.1315	&	16:40:07.395	&	$-$46:25:06.16	&	49.9	&	116.0	&	$12.4\times4.5$	&	52	\\
G338.3739$-$00.1519	&	16:41:31.132	&	$-$46:34:30.97	&	69.4	&	79.3	&	$3.8\times2.1$	&	$-$36	\\
G338.4050$-$00.2033	&	16:41:51.724	&	$-$46:35:08.44	&	212.9	&	881.0	&	$18.1\times10.9$	&	$-$20	\\
G338.4357+00.0591	&	16:40:50.351	&	$-$46:23:24.24	&	233.4	&	381.8	&	$6.7\times5.4$	&	52	\\
G338.6811$-$00.0844	&	16:42:24.015	&	$-$46:18:00.21	&	63.2	&	70.9	&	$3.0\times2.5$	&	$-$34	\\
G338.9173+00.3824	&	16:41:16.182	&	$-$45:48:53.23	&	76.9	&	82.6	&	$2.1\times1.8$	&	90	\\
G338.9217+00.6233	&	16:40:15.403	&	$-$45:39:02.93	&	48.4	&	70.3	&	$6.7\times3.2$	&	$-$53	\\
G339.1052+00.1490	&	16:42:59.560	&	$-$45:49:39.91	&	47.7	&	98.5	&	$9.9\times6.7$	&	16	\\
G339.9797$-$00.5391	&	16:49:14.767	&	$-$45:36:31.85	&	19.2	&	20.6	&	$2.5\times0.6$	&	$-$76	\\
G340.2480$-$00.3725	&	16:49:29.918	&	$-$45:17:45.07	&	38.3	&	41.3	&	$2.2\times1.9$	&	$-$76	\\
G340.2490$-$00.0460	&	16:48:05.098	&	$-$45:05:09.85	&	74.8	&	81.2	&	$2.4\times1.6$	&	72	\\
G344.2207$-$00.5953	&	17:04:13.137	&	$-$42:19:53.00	&	303.3	&	745.8	&	$11.0\times8.5$	&	50	\\
G344.4257+00.0451	&	17:02:09.575	&	$-$41:46:44.47	&	460.2	&	1336.0	&	$14.2\times9.4$	&	10	\\
G345.0034$-$00.2240	&	17:05:11.168	&	$-$41:29:04.84	&	30.5	&	33.5	&	$3.1\times2.0$	&	$-$25	\\
G345.4881+00.3148	&	17:04:28.006	&	$-$40:46:20.97	&	448.8	&	790.1	&	$8.2\times6.4$	&	7	\\
G345.5472$-$00.0801	&	17:06:19.353	&	$-$40:57:52.97	&	36.3	&	34.6	&	$0.0\times0.0$	&	0	\\
G345.6495+00.0084	&	17:06:16.186	&	$-$40:49:47.02	&	495.0	&	1251.0	&	$11.6\times8.6$	&	$-$49	\\
G346.5235+00.0839	&	17:08:42.815	&	$-$40:05:10.05	&	165.6	&	665.5	&	$15.1\times13.4$	&	42	\\
G347.2326+01.2633	&	17:06:01.947	&	$-$38:48:35.28	&	54.9	&	60.9	&	$2.9\times2.4$	&	$-$22	\\
G347.5998+00.2442	&	17:11:22.102	&	$-$39:07:26.46	&	38.2	&	43.1	&	$3.6\times2.4$	&	51	\\
G348.5312$-$00.9714	&	17:19:15.101	&	$-$39:04:33.06	&	62.9	&	166.6	&	$12.5\times8.7$	&	$-$22	\\
G348.6972$-$01.0263	&	17:19:58.889	&	$-$38:58:14.98	&	183.1	&	202.1	&	$0.0\times0.0$	&	0	\\
G348.7250$-$01.0435	&	17:20:07.076	&	$-$38:57:24.76	&	373.8	&	1139.0	&	$20.0\times5.4$	&	$-$84	\\
G348.8922$-$00.1787	&	17:16:59.926	&	$-$38:19:22.90	&	119.9	&	231.7	&	$9.8\times5.8$	&	$-$27	\\
G349.1055+00.1121	&	17:16:24.891	&	$-$37:58:50.45	&	30.5	&	34.3	&	$4.0\times2.0$	&	2	\\
\hline\\
\end{tabular}\\

\end{minipage}
\end{center}
\end{table}
\setlength{\tabcolsep}{6pt}

Spectral cubes were subsequently created of $1.5\arcmin \times 1.5\arcmin$ regions around each continuum source using the same parameters used for the continuum images, however, only a few hundred cleaning components were used to CLEAN each spectral channel. Finally, we produced an \HI\ spectrum for each continuum source by spatially averaging the \HI\ data within the 10\,per~cent contour of the emission region. Due to their minimum baselines interferometric observations effectively filter out emission from large angular scales. The shortest baseline for these observations was 337\,m and in snapshot mode are only sensitive to angular scales $\leq$1\arcmin, and thus, no background subtraction is necessary for these data.

\begin{figure*}
\begin{center}
\includegraphics[width=0.49\textwidth, trim= 0 0 0 0]{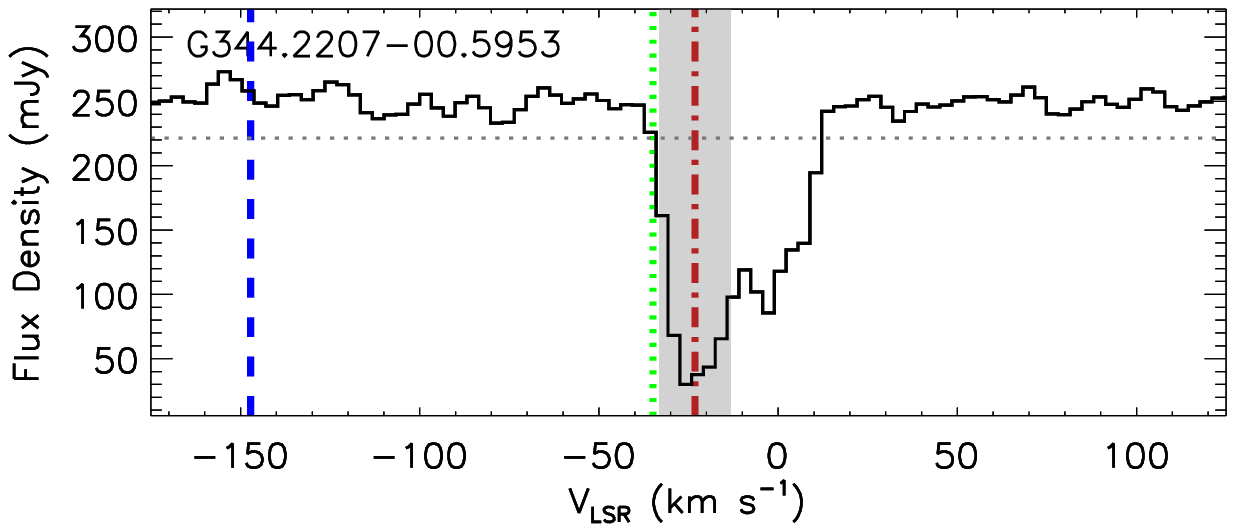}
\includegraphics[width=0.49\textwidth, trim= 0 0 0 0]{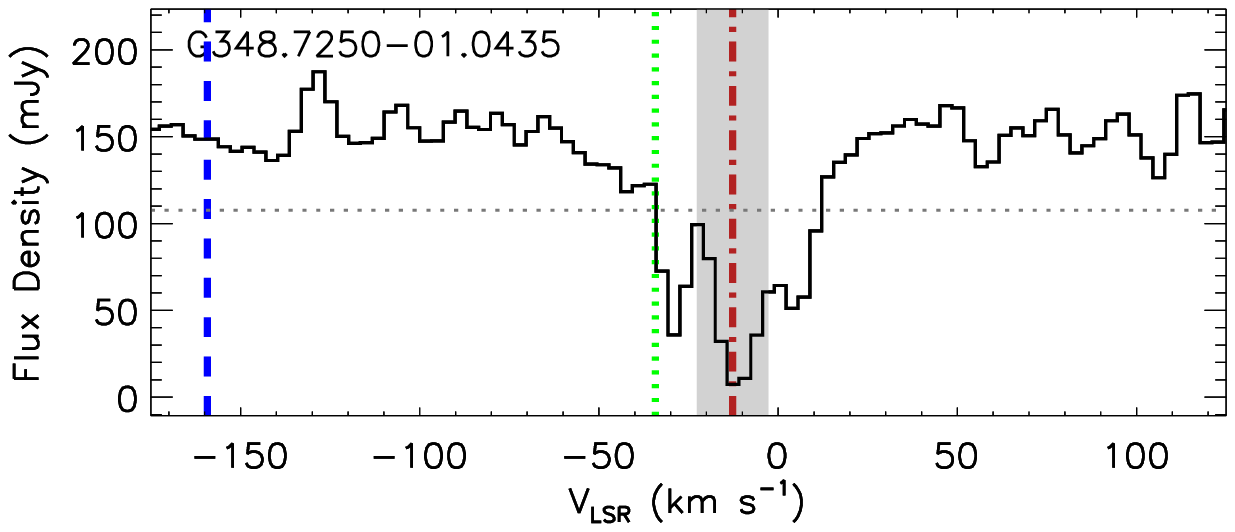}
\includegraphics[width=0.49\textwidth, trim= 0 0 0 0]{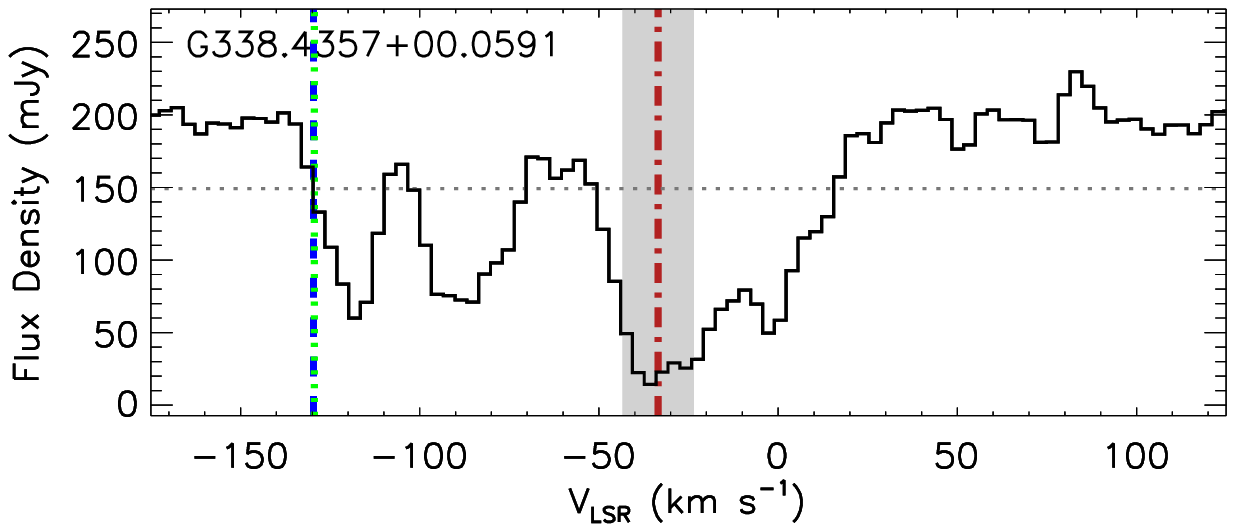}
\includegraphics[width=0.49\textwidth, trim= 0 0 0 0]{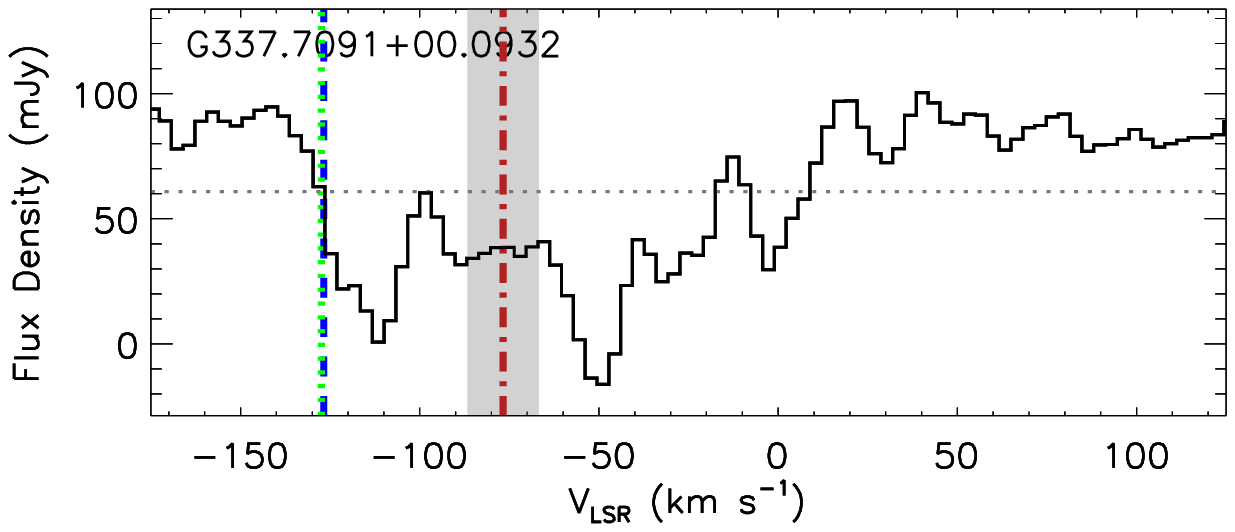}

\caption{\label{fig:c1772_spectral_data} Source-averaged, high
  resolution 
  continuum-included \HI\ spectra towards the \HII~regions
  observed with ATCA. The source
  velocity ($v_{\rm{s}}$), the velocity of the tangent point ($v_{\rm{t}}$) and the position of the
  first absorption minimum ($v_{\rm{a}}$) are shown by the red, blue and green
  vertical lines, respectively. The grey vertical band covers the velocity region 10\,\kms\ either side of the source velocity and is provided to give an indication of the uncertainty associated with it due to streaming motions. The dotted horizontal line shows the 4$\sigma_{\rm{r.m.s}}$ noise level determined from absorption free parts of the spectra (see Sect.\,\ref{sect:noise} for details). In the top and bottom panels we provide examples of sources placed at the near and far distances, respectively. The full version of this figure is
  only available in the online version of this journal.}

\end{center}
\end{figure*}

From inspection of the \HI\ spectra we set three criteria the data needed
to satisfy before the spectra were deemed usable for resolving
kinematic distance ambiguities: 1) dips in the \HI\ spectra are only
considered significant if they fall more than 4$\sigma_{\rm{r.m.s}}$ below the mean
spectral continuum level as determined from absorption free regions of
the spectra; 2) the continuum level is greater than 30\,mJy
beam$^{-1}$ channel$^{-1}$ to ensure a 4$\sigma_{\rm{r.m.s}}$ detection of the
continuum and 3) absorption must be present at the
same velocity as the \HII~region's host cloud to ensure the source is
genuinely associated with cold interstellar gas. The last of these criteria stems from the assumption that the HII region is still deeply embedded within its natal molecular cloud, which should produce an absorption at a similar velocity in the 21\,cm spectrum to that of the molecular line velocity ($\pm$10\,\kms); if this is not the case it may be that the molecular cloud and radio emission have been incorrectly associated. In total the \HI\ data towards 53 \HII~regions satisfied these selection criteria; a
sample of the continuum maps and \HI\ spectra are presented in
Figs.\,\ref{fig:example_images} and \ref{fig:c1772_spectral_data},
respectively.

\subsection{Archival Data Sets: SGPS and VGPS data}
\label{sect:sgps}

We complement our targeted high-resolution observations with lower-resolution, \HI\ continuum-included data extracted from the Southern Galactic Plane Survey (SGPS; \citealt{mcclure2005}) and the VLA Galactic Plane Survey \citep[VGPS]{stil2006}. The SGPS combines interferometric observations conducted with the Compact Array and Parkes single-dish data to cover two regions; SGPS\,I \emph{l}=253-358\degr\ and SPGS\,II \emph{l}=5-20\degr\ with a Galactic latitude coverage of $|b| \leq$  1.5\degr, an angular resolution of $\sim$2\arcmin, and an r.m.s. sensitivity of $\sim$1\,K. The VGPS covers the range of Galactic longitudes from $18^{\circ}$ to $67^{\circ}$ in the Galactic first quadrant. The latitude coverage increases with longitude from $|b| < 1.3\degr$ to $|b| < 2.3\degr$. 21\,cm data was taken using the VLA and the Green Bank Telescope which were subsequently combined to produce data cubes with 1\arcmin$\times$1\arcmin$\times$1.56\,\kms\ resolution, with velocity channels of 0.824\,\kms.  

To identify RMS sources associated with 21\,cm radio emission we
extracted continuum emission maps from the archives and searched for
positional coincidences between the RMS and radio sources. Continuum-included \HI\ spectra were subsequently extracted towards all RMS
sources found to be associated with 21\,cm emission, spatially
integrating over the radio emission region to obtain a  source averaged
\HI\ spectrum towards each source. Unlike the targeted observations
discussed in the previous section these data include short spacing
information and therefore include a contribution from the large-scale
background emission that has been removed from the source-averaged
spectra. 

\begin{figure*}
\begin{center}
\includegraphics[width=0.49\textwidth, trim= 0 0 0 0]{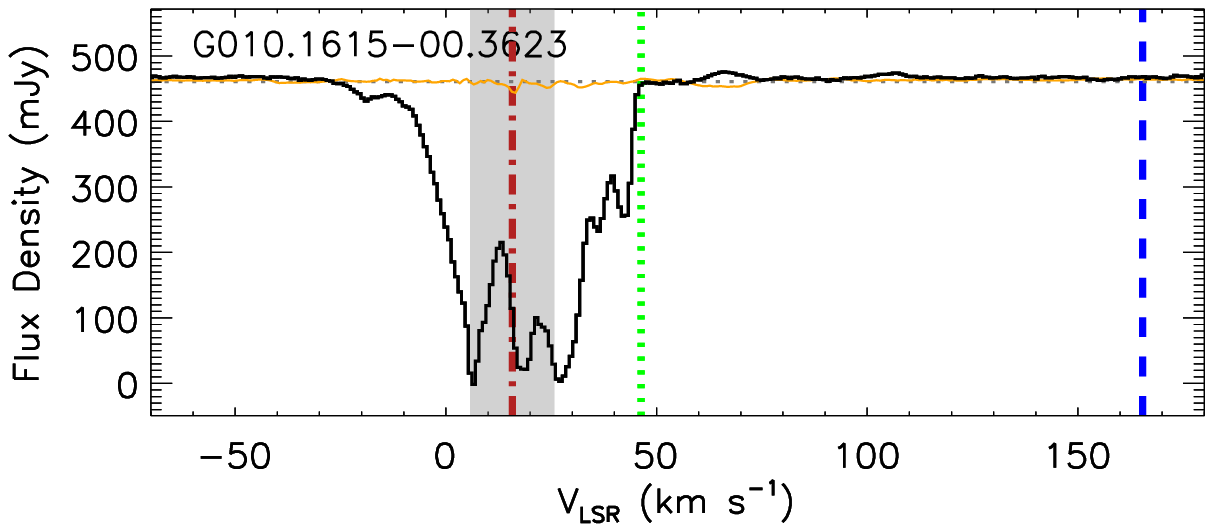}
\includegraphics[width=0.49\textwidth, trim= 0 0 0 0]{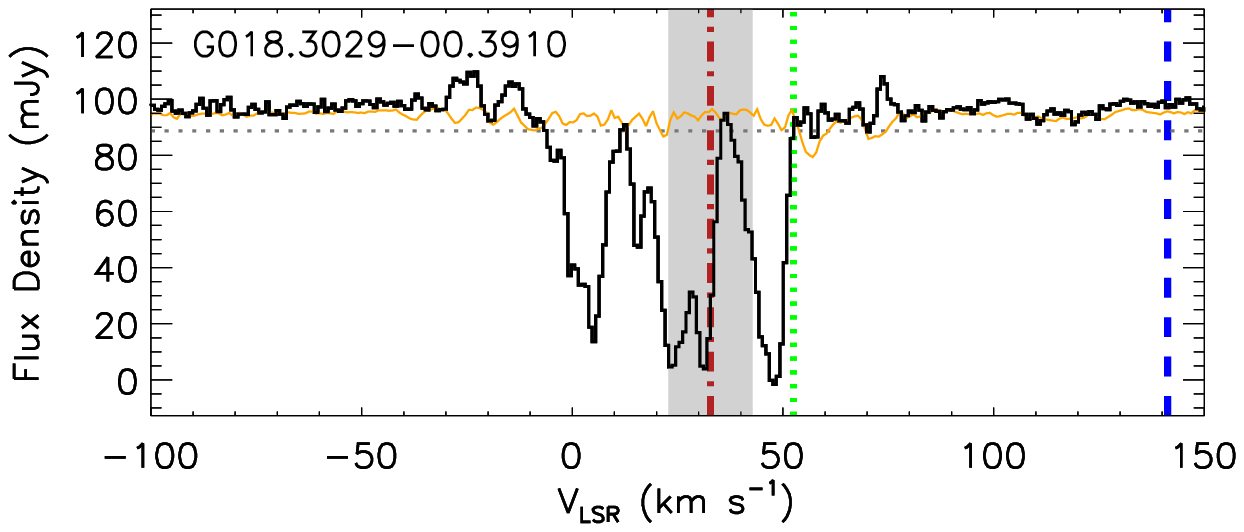}
\includegraphics[width=0.49\textwidth, trim= 0 0 0 0]{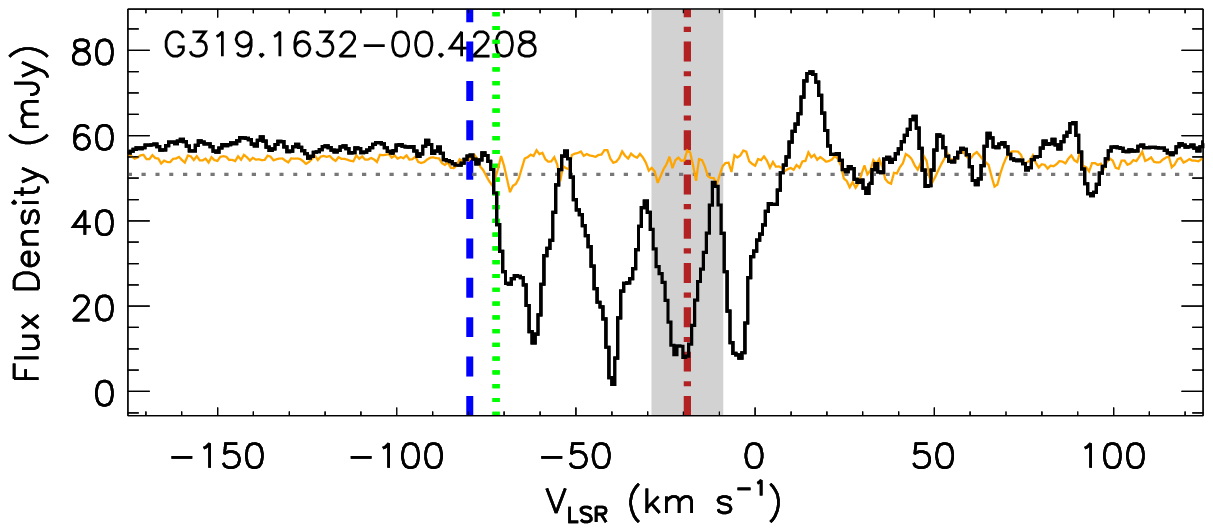}
\includegraphics[width=0.49\textwidth, trim= 0 0 0 0]{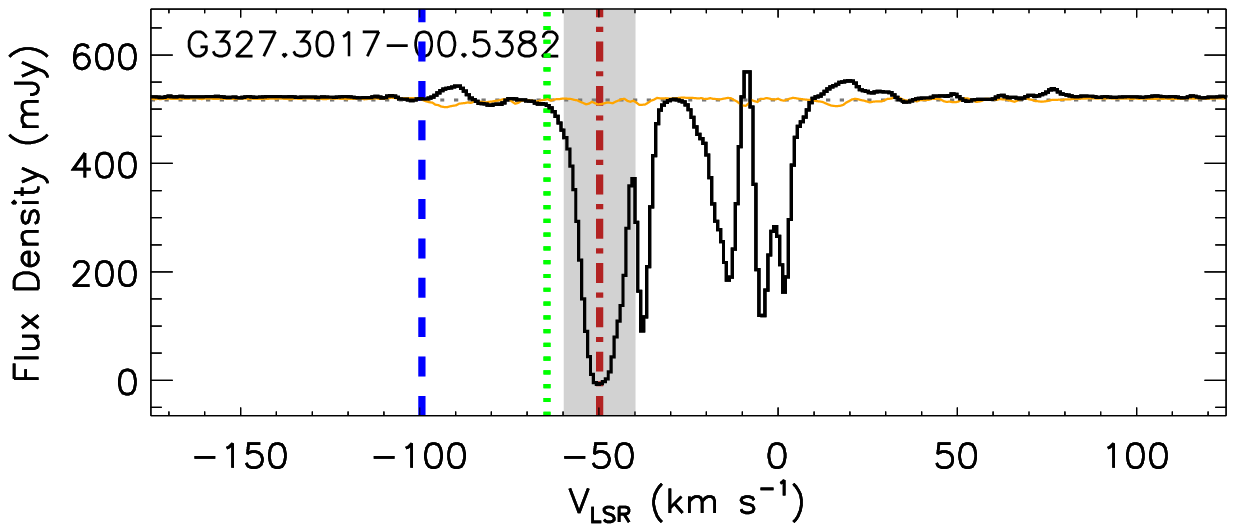}

\caption{\label{fig:sgps_spectral_data} Source-averaged low resolution
  \HI\ spectra towards the \HII~regions extracted from \HI\
  SGPS and VGPS survey data. The source
  velocity ($v_{\rm{s}}$), the velocity of the tangent point ($v_{\rm{t}}$) and the position of the
  first absorption minimum ($v_{\rm{a}}$) are shown by the red, blue and green
  vertical lines, respectively. The grey vertical band covers the velocity region 10\,\kms\ either side of the source velocity and is provided to give an indication of the uncertainty associated with it due to streaming motions. The dotted horizontal and solid yellow lines shows the 4$\sigma_{\rm{r.m.s}}$ receiver noise level determined from absorption free parts of the spectra and the \HI\ emission fluctuations, respectively (see Sect.\,\ref{sect:noise} for details). In the top and bottom panels we provide examples of sources placed at the near and far distances, respectively. The full version of this figure is only
  available in the online version of this journal.}

\end{center}
\end{figure*}

Following \citet{anderson2009a} the background contribution was estimated by averaging the emission found within four regions located as near as possible to the continuum source taking care to avoid any other nearby continuum sources. These regions were chosen to surround the target source to compensate for any gradients that may be present in the background emission. The source averaged spectra were determined from a small region centred on the strongest continuum emission. The final spectra were produced by subtracting the background emission from the source-averaged spectra. In Fig.\,\ref{fig:sgps_spectral_data} we present a sample of the background subtracted \HI\ spectra obtained towards RMS sources from the SGPS and the VGPS.

\subsection{Receiver noise and emission fluctuation}
\label{sect:noise}

There are two important sources of noise that need to be considered when determining the reliability of any particular absorption feature; these are the receiver noise and \HI\ emission fluctuations. We have estimated the receiver noise by calculating the standard deviation using the absorption free channels in each spectra. We set a threshold value to be 4 times the standard deviation (4$\sigma_{\rm{r.m.s}}$) and require the absorption to be larger than this to be considered significant. This threshold value is shown on the plots presented in Figs.\,3 and 4 by a dashed horizontal line.

The second source of noise is due to \HI\ emission fluctuations. These are present on all angular scales (\citealt{green1993}) and can result in positive and negative wiggles in the observed spectrum, particularly where the \HI\ emission is bright, which can confuse the analysis. To estimate the emission fluctuations in the background we have calculated the standard deviation of the off-source spectra as a function of velocity (cf. \citealt{anderson2009a}). The emission fluctuations are a strong function of baseline length and therefore we might expect them to be lower for our targeted observations than for the archival SGPS and VGPS data, and this is indeed the case. In fact we found the emission fluctuations for the targeted observations to be much smaller than the threshold value derived from the receiver noise and so can be neglected. However, the emission fluctuations estimated for the archival data are in many cases comparable to the receiver noise and so need to be considered; these are shown on the plots presented in Fig.\,4 by the yellow horizontal line.  In these cases, since the emission fluctuations are not independent of the receiver noise, we require that the depth of the absorption feature must be larger than 4 times the receiver noise (i.e., $\sigma_{\rm{r.m.s}}$) and larger than the background emission fluctuations present in the spectrum to be considered reliable.  

\section{Results and analysis}
\label{sect:results}

In total we have identified 122 \HII~regions with strong continuum
emission displaying significant \HI\ absorption --- 53 high-resolution, ATCA-targeted observations and 69 drawn from the SGPS and VGPS data sets. The two data
sets have 17 sources in common. The majority of the \HII~regions targeted in these observations were too weak to provide a significant continuum in the lower resolution survey data that are instead dominated by \HI\ emission and therefore do not satisfy our criteria for inclusion here. Therefore the total number of
discrete sources in the combined data sets is 105. In this section we
describe the method used to resolve the distance ambiguities towards
these \HII~regions. 

We have calculated distances for all sources that have been assigned a unique kinematic velocity using the \citet{brand1993} Galactic rotation curve and assuming the distance to the Galactic centre to be 8.5\,kpc and the radial velocity at the position of the Sun to be 220\,\kms. In Table\,\ref{tbl:analysis_results} we present the source names, velocity, kinematic distances, assigned distance and bolometric luminosity --- the distance assignments will be discussed in the next section. In the following two subsections we test the reliability of these results by comparing the distance solutions obtained for the sources for which both high- and low-resolution data are available and with distances assigned by previous studies reported in the literature (see last column of Table\,\ref{tbl:analysis_results} for references).

\subsection{Resolving the Kinematic Distance Ambiguity}

We have based our kinematic distance ambiguity resolutions for this sample of \HII~regions on work presented by \citet{kolpak2003}. These authors measured the source velocity ($v_{\rm{S}}$), the velocity of the tangent point ($v_{\rm{T}}$) and the velocity of the highest velocity of absorption ($v_{\rm{A}}$) of a sample of Northern hemisphere \HII~regions and found their sources fell into two distinct groups; those where the $v_{\rm{T}}$-$v_{\rm{S}}\simeq 0$ and showed no dependence on $v_{\rm{T}}$-$v_{\rm{A}}$, and those where $v_{\rm{T}}$-$v_{\rm{S}}$ increases with increasing $v_{\rm{T}}$-$v_{\rm{A}}$. These two groups are consistent with expectations for \HII~regions located at the far and near distance respectively. 

Using the \citet{kolpak2003} method requires only measuring the three
velocities. Making the assumption that the \HII~region is still
embedded within its natal molecular cloud, we can use $^{13}$CO
emission to determine the source velocity (i.e.,
\citealt{urquhart_13co_south,urquhart_13co_north}). We determine the
velocity of the tangent point using the empirical relationship between
the \HI\ termination velocities and Galactic longitude derived by
\citet{mcclure2005}. Finally, we measure the absorption velocity by
estimating the standard deviation of the \HI\ spectrum from absorption
free channels, and measuring the minimum velocity where the absorption
dips is larger than 4$\sigma_{\rm{r.m.s}}$ and larger than the value of \HI\ emission fluctuations --- for sources located in
the Northern Galactic Plane this becomes the maximum velocity. In
Table\,\ref{tbl:analysis_results} we present the measured velocity components for each \HII~region.

In Figs.\,\ref{fig:c1772_spectral_data} and
\ref{fig:sgps_spectral_data} we present plots of the 21\,cm continuum spectrum
towards each source where significant \HI\ absorption is seen. On these
plots the velocity of the tangent point, the source velocity and the
absorption velocity are indicated by the blue, red and green vertical
lines, respectively. The grey horizontal and yellow line show the 4$\sigma_{\rm{r.m.s}}$
threshold level determined from the receiver noise and the \HI\ emission fluctuations below which absorption is considered significant (see Sect.\,\ref{sect:noise} for details). The grey
shaded region shows the possible deviation of the source velocity from
pure circular rotation due to streaming motions
($\pm10\,{\rm{km}}\,{\rm{s}}^{-1}$; \citealt{burton1971,
  stark1989}). Of these we find three where the source velocity is
within $10\,{\rm{km}}\,{\rm{s}}^{-1}$ of the tangent velocity and
since this is smaller than the uncertainly introduced by the streaming
motions, we have placed these sources at the distance of the tangent.

\begin{figure}
\includegraphics[width=0.95\linewidth]{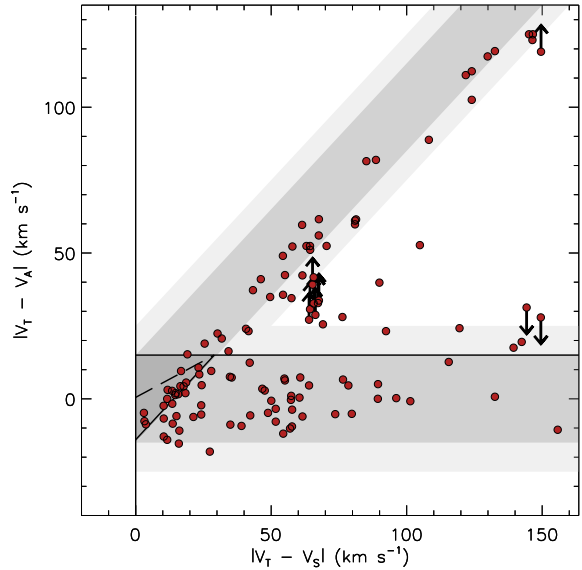}

\caption{\label{fig:kolpak} Plot of the differences in velocity
  between the tangent point, the RMS source, and the velocity of the
  absorption as defined in the text. The diagonal and horizontal
  shaded regions show the expected locations of sources at the near
  and far distances respectively, while the solid lines mark the
  approximate boundaries.} 
\end{figure}

To determine the correct distance assignments Kolpak et al. produced a
simulation with 10,000 \HII~regions the results of which are included
in Fig.\,\ref{fig:kolpak}; the diagonal and horizontal shaded regions
show the expected locations in the velocity parameter space of sources
at the near and far distances, respectively. The dark grey bands mark
the region of parameter space where 90\,per\,cent of the simulated
sources were located, and the lighter grey region outlines a
10\,\kms\ extension to the simulated data to allow for streaming
motions. Sources located in the diagonal shaded region of the plot can
be assigned a near distance with a high degree of confidence, and
sources located in the horizontal shaded region of the plot can be
assigned a far distance with a high degree of confidence. The darker
triangular region located towards the lower left quadrant of the plot
indicates the overlapping region of parameter space where the distance
assignments are inherently more uncertain. This region is divided by a
dashed line with sources located above the line being assigned to the
near distance, and sources located below being assigned a far
distance, however, both with a lower degree of confidence. 

Using the plot presented in Fig.\,\ref{fig:kolpak} we are able to
assign distances to 95 \HII~regions depending on their association with
the various regions discussed in the previous paragraph. We place 59 at the far distance, 33 at the near distance and three at the tangent. In Table\,\ref{tbl:analysis_results} we present derived distances and distance assignments; we denoted lower confidence assignments by appending a `?' to the distance assignment. 

We find 10 sources that are located between the grey horizontal and diagonal bands of the plot; we indicate these sources by placing an ellipsis in the KDS column of Table\,\ref{tbl:analysis_results}. The available velocity data are not sufficient to resolve the distance ambiguity for these sources, and
additional information has been sought before a distance can been
assigned. In the following subsection we will discuss the results of a literature search which has allowed us to resolve
the distances to eight of these sources, two at the far distance and six at the near distance. The distance allocations for these eight sources are indicated on the plot presented in Fig.\,\ref{fig:kolpak} by the up
and down arrows, respectively. We were unable to find any additional information for the remaining two sources (i.e., G338.9173+00.3824, G340.2768$-$00.2104) and consequently no distance has been assigned.

\setlength{\tabcolsep}{4pt}
\begin{table*}

\begin{center}
\caption{Summary of measured source, tangent and absorption velocities, kinematic distances and results of our kinematic distance ambiguity analysis.}
\label{tbl:analysis_results}
\begin{minipage}{\linewidth}
\scriptsize
\begin{tabular}{lcc...c...cc..cc}

\hline \hline
 & &	  		  		& \multicolumn{3}{c}{Measured \vlsr}& & \multicolumn{3}{c}{Rotation Model} && \multicolumn{4}{c}{Results}   \\ 
\cline{4-6} \cline{8-10}\cline{12-15}
MSX Name&	 	RA  		&Dec & \multicolumn{1}{c}{$v_{\rm{S}}$} &  \multicolumn{1}{c}{$v_{\rm{T}}$} &  \multicolumn{1}{c}{$v_{\rm{A}}$} &&  \multicolumn{1}{c}{Near} & \multicolumn{1}{c}{Far} &  \multicolumn{1}{c}{RGC} && KDA &  \multicolumn{1}{c}{Distance}  & \multicolumn{1}{c}{z} &  \multicolumn{1}{c}{Log(Lum)} & Reference\\
&	(J2000) 	& (J2000) 	&  \multicolumn{1}{c}{(\kms)} & \multicolumn{1}{c}{(\kms)} & \multicolumn{1}{c}{(\kms)} &&   \multicolumn{1}{c}{(kpc)} &  \multicolumn{1}{c}{(kpc)} & \multicolumn{1}{c}{(kpc)} && solution & \multicolumn{1}{c}{(kpc)} & \multicolumn{1}{c}{(pc)} & \multicolumn{1}{c}{(\lsun)}  & \\
\hline
G010.1615$-$00.3623$^\dagger$	&	18:09:26.88	&	$-$20:19:28.2	&	15.8	&	165.3	&	46.3	&&	2.5	&	14.2	&	6.1	& &	\multicolumn{1}{c}{$\cdots$}	&	\multicolumn{1}{c}{{\bf 3.55$^{4}$}}	&	-22.4	&	5.4	&	1, 2, 3, 4	\\
G012.8062$-$00.1987$^\dagger$	&	18:14:13.55	&	$-$17:55:37.5	&	34.4	&	156.2	&	45.2	&&	3.7	&	12.9	&	5.0	& &	n	&	3.7	&	-12.9	&	5.5	&	7	\\
G015.0357$-$00.6795$^\dagger$	&	18:20:25.51	&	$-$16:11:35.5	&	20.3	&	150.2	&	32.8	&&	2.3	&	14.1	&	6.3	& &	n	&	\multicolumn{1}{c}{1.98$^{5}$}	&	-23.5	&	5.1	&	4, 5	\\
G018.3029$-$00.3910$^\dagger$	&	18:25:42.48	&	$-$13:10:20.2	&	33.1	&	141.3	&	52.5	&&	3.0	&	13.1	&	5.7	& &	n	&	3.0	&	-20.4	&	4.6	&	6,8,14	\\
G019.7403+00.2799$^\dagger$	&	18:26:01.48	&	$-$11:35:16.4	&	19.9	&	135.4	&	122.8	&&	1.9	&	14.1	&	6.7	& &	f	&	14.1	&	68.8	&	4.9	&		\\
G028.3046$-$00.3871$^\dagger$	&	18:44:21.91	&	$-$04:17:39.1	&	84.9	&	112.8	&	103.2	&&	5.2	&	9.8	&	4.6	& &	f	&	9.8	&	-35.0	&	4.6	&	7	\\
G032.2718$-$00.2260$^\dagger$	&	18:51:02.32	&	$-$00:41:26.1	&	22.3	&	102.0	&	107.2	&&	1.6	&	12.8	&	7.2	& &	f	&	12.8	&	-50.4	&	4.9	&		\\
G033.8104$-$00.1869$^\dagger$	&	18:53:42.38	&	+00:41:47.7	&	42.4	&	96.8	&	108.8	&&	2.9	&	11.3	&	6.3	& &	f	&	11.3	&	-36.8	&	5.3	&	8, 14↵	\\
G039.7279$-$00.3974$^\dagger$	&	19:05:18.00	&	+05:51:47.1	&	57.8	&	82.0	&	87.4	&&	3.9	&	9.2	&	6.0	& &	f	&	9.2	&	-63.5	&	4.5	&		\\
G039.8821$-$00.3457$^\dagger$	&	19:05:24.02	&	+06:01:25.6	&	57.8	&	82.0	&	84.2	&&	3.9	&	9.1	&	6.0	& &	f	&	9.1	&	-55.0	&	4.7	&		\\
G052.7528+00.3343$^\dagger$	&	19:27:32.30	&	+17:43:27.1	&	15.2	&	54.3	&	63.6	&&	1.0	&	9.2	&	7.9	& &	f	&	9.2	&	54.0	&	4.5	&	14	\\
G056.3694$-$00.6333$^\dagger$	&	19:38:31.56	&	+20:25:18.8	&	32.4	&	48.9	&	44.6	&&	3.0	&	6.4	&	7.3	& &	f?	&	6.4	&	-71.1	&	3.8	&		\\
G060.8828$-$00.1295$^\dagger$	&	19:46:20.18	&	+24:35:23.2	&	21.7	&	40.8	&	25.5	&&	2.0	&	6.3	&	7.7	& &	n	&	2.0	&	-4.6	&	4.3	&		\\
G063.1720+00.4425$^\dagger$	&	19:49:16.54	&	+26:51:16.9	&	21.8	&	37.9	&	48.8	&&	2.3	&	5.4	&	7.7	& &	f	&	5.4	&	41.3	&	3.5	&		\\
G281.5576$-$02.4775	&	09:58:02.85	&	$-$57:57:48.9	&	-7.0	&	-21.8	&	-19.8	&&	1.1	&	2.3	&	8.3	& &	f?	&	2.3	&	-98.3	&	4.4	&		\\
G281.8449$-$01.6094	&	10:03:40.96	&	$-$57:26:39.8	&	-7.2	&	-21.8	&	-20.5	&&	1.1	&	2.4	&	8.3	& &	f?	&	2.4	&	-66.6	&	4.0	&		\\
G305.1967+00.0335	&	13:11:14.61	&	$-$62:45:04.3	&	-35.6	&	-50.6	&	-56.6	&&	3.2	&	6.6	&	7.2	& &	f	&	6.6	&	3.9	&	$<$5.8	&		\\
G305.2694$-$00.0072	&	13:11:54.31	&	$-$62:47:10.3	&	-32.3	&	-50.6	&	-48.7	&&	2.8	&	7.1	&	7.3	& &	f	&	7.1	&	-0.9	&	5.2	&		\\
G305.3500+00.2240	&	13:12:26.56	&	$-$62:32:57.1	&	-39.0	&	-50.6	&	-50.7	&&	3.7	&	6.1	&	7.0	& &	f?	&	6.1	&	23.9	&	4.6	&	9	\\
G307.5606$-$00.5871	&	13:32:31.15	&	$-$63:05:21.1	&	-33.0	&	-54.3	&	-60.5	&&	2.6	&	7.7	&	7.2	& &	f	&	7.7	&	-79.3	&	5.1	&		\\
G307.6138$-$00.2559	&	13:32:31.10	&	$-$62:45:13.6	&	-36.6	&	-54.3	&	-50.0	&&	3.0	&	7.4	&	7.1	& &	f?	&	7.4	&	-32.9	&	4.8	&		\\
G307.6213$-$00.2622	&	13:32:35.49	&	$-$62:45:31.3	&	-37.6	&	-54.3	&	-44.8	&&	3.1	&	7.3	&	7.0	& &	f?	&	7.3	&	-33.3	&	4.8	&		\\
G308.6542+00.6039$^\dagger$	&	13:40:02.44	&	$-$61:43:36.1	&	-47.8	&	-58.1	&	-60.5	&&	4.5	&	6.1	&	6.7	& &	f	&	6.1	&	64.4	&	3.7	&	10	\\
G311.4255+00.5964	&	14:02:36.38	&	$-$61:05:46.6	&	-48.6	&	-62.1	&	-63.8	&&	3.9	&	7.3	&	6.6	& &	f	&	7.3	&	76.4	&	4.5	&		\\
G311.6380+00.3009$^\dagger$	&	14:04:58.84	&	$-$61:19:17.7	&	-48.2	&	-64.2	&	-79.5	&&	3.8	&	7.5	&	6.6	& &	f	&	7.5	&	39.2	&	3.4	&	10↵	\\
G313.4573+00.1934$^\dagger$	&	14:19:34.89	&	$-$60:51:51.4	&	-4.7	&	-66.2	&	-72.3	&&	0.2	&	11.5	&	8.4	& &	f	&	11.5	&	38.9	&	5.2	&	10↵	\\
G316.1386$-$00.5009	&	14:42:01.58	&	$-$60:30:20.1	&	-61.0	&	-72.8	&	-69.7	&&	4.6	&	7.6	&	6.1	& &	f?	&	7.6	&	-66.9	&	3.7	&		\\
G316.7754$-$00.0447$^\dagger$	&	14:45:08.54	&	$-$59:49:29.2	&	-39.6	&	-75.0	&	-67.7	&&	2.7	&	9.7	&	6.8	& &	f	&	9.7	&	-7.5	&	5.1	&	11	\\
G317.4112+00.1050$^\dagger$	&	14:49:10.27	&	$-$59:24:56.8	&	-40.8	&	-75.0	&	-58.7	&&	2.8	&	9.7	&	6.7	& &	n	&	2.8	&	5.1	&	4.3	&		\\
G318.7251$-$00.2241	&	14:59:30.09	&	$-$59:06:41.7	&	-22.4	&	-77.3	&	-70.3	&&	1.5	&	11.3	&	7.4	& &	f	&	11.3	&	-44.1	&	4.7	&		\\
G318.7748$-$00.1513	&	14:59:34.53	&	$-$59:01:26.0	&	-36.6	&	-77.3	&	-53.3	&&	2.5	&	10.3	&	6.8	& &	n	&	2.5	&	-6.6	&	3.8	&		\\
G318.9148$-$00.1647	&	15:00:34.94	&	$-$58:58:10.2	&	-22.2	&	-77.3	&	-71.0	&&	1.5	&	11.3	&	7.4	& &	f	&	11.3	&	-32.5	&	5.5	&	10, 11↵	\\
G319.1632$-$00.4208$^\dagger$	&	15:03:13.84	&	$-$59:04:30.0	&	-18.9	&	-79.6	&	-72.3	&&	1.3	&	11.6	&	7.6	& &	f	&	11.6	&	-85.2	&	5.5	&		\\
G319.3622+00.0126$^\dagger$	&	15:02:57.40	&	$-$58:35:57.8	&	-19.2	&	-79.6	&	-79.2	&&	1.3	&	11.6	&	7.6	& &	f	&	11.6	&	2.6	&	4.7	&	13	\\
G320.1750+00.8001$^\dagger$	&	15:05:25.36	&	$-$57:30:56.1	&	-39.9	&	-82.0	&	-69.6	&&	2.7	&	10.4	&	6.7	& &	f	&	10.4	&	144.8	&	4.6	&	13	\\
G320.2434$-$00.2801$^\dagger$	&	15:09:55.84	&	$-$58:25:05.1	&	-68.3	&	-82.0	&	-90.5	&&	4.8	&	8.3	&	5.7	& &	f	&	8.3	&	-40.6	&	4.4	&		\\
G321.0523$-$00.5070$^\dagger$	&	15:16:05.90	&	$-$58:11:43.8	&	-60.9	&	-84.4	&	-76.0	&&	4.1	&	9.1	&	5.9	& &	f	&	9.1	&	-80.8	&	4.9	&	13	\\
G322.1729+00.6442$^\dagger$	&	15:18:38.13	&	$-$56:37:32.5	&	-56.6	&	-86.8	&	-64.4	&&	3.8	&	9.7	&	6.0	& &	n	&	3.8	&	42.2	&	4.0	&	9	\\
G324.1997+00.1192$^\dagger$	&	15:32:53.13	&	$-$55:56:14.2	&	-88.0	&	-91.8	&	-100.5	&&	6.2	&	7.6	&	5.0	& &	tp	&	6.9	&	14.3	&	5.7	&		\\
G326.4719$-$00.3777	&	15:47:49.80	&	$-$54:58:34.3	&	-55.2	&	-96.8	&	-73.6	&&	3.6	&	10.6	&	5.8	& &	n	&	3.6	&	-23.8	&	4.6	&		\\
G326.7249+00.6159	&	15:44:59.44	&	$-$54:02:13.9	&	-42.5	&	-96.8	&	-61.2	&&	2.9	&	11.4	&	6.3	& &	n	&	\multicolumn{1}{c}{1.82$^{4}$}	&	19.6	&	3.6	&	4, 13	\\
G327.3017$-$00.5382$^\dagger$	&	15:53:00.76	&	$-$54:34:53.0	&	-49.8	&	-99.4	&	-64.5	&&	3.3	&	11.0	&	6.0	& &	n	&	3.3	&	-31.0	&	4.8	&	13	\\
G327.7579$-$00.3515$^\dagger$	&	15:54:36.88	&	$-$54:08:49.9	&	-76.6	&	-102.0	&	-83.1	&&	4.8	&	9.6	&	5.1	& &	n	&	4.8	&	-29.6	&	4.7	&		\\
G327.8483+00.0175	&	15:53:29.47	&	$-$53:48:18.0	&	-51.8	&	-99.4	&	-96.6	&&	3.4	&	11.0	&	5.9	& &	f	&	11.0	&	3.4	&	4.4	&		\\
G328.3067+00.4308	&	15:54:06.23	&	$-$53:11:40.2	&	-91.7	&	-102.0	&	-115.0	&&	5.8	&	8.7	&	4.7	& &	f	&	\multicolumn{1}{c}{5.80$^{4}$}	&	43.6	&	5.4	&	4	\\
G328.5739$-$00.5483$^\dagger$	&	15:59:43.08	&	$-$53:46:17.0	&	-47.4	&	-104.7	&	-105.0	&&	3.2	&	11.3	&	6.0	& &	f	&	11.3	&	-108.5	&	5.4	&	13	\\
G328.5759$-$00.5285$^\dagger$	&	15:59:38.44	&	$-$53:45:18.3	&	-47.0	&	-104.7	&	-108.4	&&	3.2	&	11.4	&	6.0	& &	f	&	11.4	&	-104.8	&	6.5	&	13	\\
G329.4720+00.2143	&	16:00:55.89	&	$-$52:36:26.2	&	-101.5	&	-104.7	&	-112.3	&&	6.4	&	8.2	&	4.4	& &	tp	&	7.3	&	27.4	&	4.8	&		\\
G330.2935$-$00.3946	&	16:07:38.06	&	$-$52:31:03.7	&	-80.0	&	-107.4	&	-125.5	&&	4.9	&	9.8	&	4.9	& &	f	&	9.8	&	-67.7	&	5.2	&		\\
G330.8708$-$00.3715$^\dagger$	&	16:10:19.00	&	$-$52:06:38.5	&	-63.3	&	-110.1	&	-106.6	&&	4.1	&	10.8	&	5.3	& &	n	&	4.1	&	-26.4	&	3.9	&	13	\\
G330.9544$-$00.1817	&	16:09:52.77	&	$-$51:54:52.2	&	-91.6	&	-107.4	&	-105.8	&&	5.6	&	9.3	&	4.5	& &	f?	&	9.3	&	-29.5	&	6.0	&	10	\\
G331.1194$-$00.4955$^\dagger$	&	16:12:03.04	&	$-$52:01:55.5	&	-66.8	&	-110.1	&	-72.8	&&	4.3	&	10.6	&	5.2	& &	n	&	4.3	&	-36.8	&	4.1	&		\\
G331.1465+00.1343	&	16:09:24.55	&	$-$51:33:06.8	&	-75.3	&	-110.1	&	-102.5	&&	4.7	&	10.2	&	4.9	& &	f	&	10.2	&	23.9	&	4.3	&		\\
G331.3546+01.0638	&	16:06:24.16	&	$-$50:43:27.4	&	-78.3	&	-110.1	&	-89.4	&&	4.8	&	10.1	&	4.8	& &	n	&	4.8	&	89.9	&	5.2	&		\\
G331.4181$-$00.3546	&	16:12:50.25	&	$-$51:43:29.9	&	-63.9	&	-110.1	&	-69.0	&&	4.1	&	10.8	&	5.3	& &	n	&	4.1	&	-25.5	&	4.3	&		\\
G331.5414$-$00.0675$^\dagger$	&	16:12:09.09	&	$-$51:25:52.6	&	-88.4	&	-112.8	&	-108.1	&&	5.4	&	9.6	&	4.6	& &	f	&	9.6	&	-11.3	&	5.6	&	13	\\
G332.1544$-$00.4487$^\dagger$	&	16:16:40.77	&	$-$51:17:06.3	&	-55.3	&	-112.8	&	-78.3	&&	3.7	&	11.3	&	5.5	& &	f	&	\multicolumn{1}{c}{3.96$^{4}$}	&	-31.0	&	5.4	&	4	\\
G332.2944$-$00.0962	&	16:15:45.86	&	$-$50:56:02.3	&	-48.4	&	-112.8	&	-61.7	&&	3.3	&	11.7	&	5.8	& &	n	&	\multicolumn{1}{c}{3.96$^{4}$}	&	-6.6	&	4.5	&	4, 9, 10, 11	\\
G332.5438$-$00.1277	&	16:17:02.47	&	$-$50:47:00.9	&	-47.5	&	-112.8	&	-73.6	&&	3.3	&	11.8	&	5.8	& &	\multicolumn{1}{c}{$\cdots$}	&	\multicolumn{1}{c}{{\bf 3.96$^{4}$}}	&	-8.8	&	5.0	&	4, 10	\\
G332.6450$-$00.6036$^\dagger$	&	16:19:36.57	&	$-$51:03:12.2	&	-51.6	&	-115.5	&	-88.4	&&	3.5	&	11.6	&	5.6	& &	\multicolumn{1}{c}{$\cdots$}	&	\multicolumn{1}{c}{{\bf 3.96$^{4}$}}	&	-41.7	&	4.4	&	4	\\
G332.8256$-$00.5498	&	16:20:11.18	&	$-$50:53:17.5	&	-57.7	&	-112.8	&	-70.3	&&	3.8	&	11.3	&	5.4	& &	n	&	\multicolumn{1}{c}{3.96$^{4}$}	&	-38.0	&	5.4	&	4	\\
G333.0058+00.7707$^\dagger$	&	16:15:13.43	&	$-$49:48:56.8	&	-48.3	&	-115.5	&	-82.7	&&	3.3	&	11.8	&	5.7	& &	\multicolumn{1}{c}{$\cdots$}	&	\multicolumn{1}{c}{{\bf 3.3}}	&	45.0	&	3.3	&	13	\\
G333.0145$-$00.4438$^\dagger$	&	16:20:33.81	&	$-$50:40:48.3	&	-54.0	&	-115.5	&	-73.2	&&	3.7	&	11.5	&	5.5	& &	n	&	\multicolumn{1}{c}{3.96$^{4}$}	&	-30.7	&	5.5	&	4	\\
G333.0162+00.7615	&	16:15:18.64	&	$-$49:48:55.0	&	-47.9	&	-115.5	&	-53.9	&&	3.3	&	11.8	&	5.7	& &	n	&	3.3	&	44.2	&	$<$4.8	&	13	\\
G333.1306$-$00.4275	&	16:21:00.64	&	$-$50:35:12.1	&	-51.2	&	-115.5	&	-63.1	&&	3.5	&	11.7	&	5.6	& &	n	&	\multicolumn{1}{c}{3.96$^{4}$}	&	-29.5	&	5.7	&	4, 13	\\
G333.2880$-$00.3907	&	16:21:32.95	&	$-$50:26:58.2	&	-52.5	&	-115.5	&	-63.1	&&	3.6	&	11.6	&	5.5	& &	n	&	\multicolumn{1}{c}{3.96$^{4}$}	&	-27.0	&	4.9	&	4	\\
G333.3072$-$00.3666	&	16:21:31.63	&	$-$50:25:08.0	&	-49.9	&	-115.5	&	-82.8	&&	3.4	&	11.8	&	5.6	& &	\multicolumn{1}{c}{$\cdots$}	&	\multicolumn{1}{c}{{\bf 3.96$^{4}$}}	&	-25.3	&	5.7	&	4	\\
G333.6032$-$00.2184	&	16:22:10.87	&	$-$50:06:17.2	&	-49.2	&	-115.5	&	-86.7	&&	3.4	&	11.8	&	5.6	& &	\multicolumn{1}{c}{$\cdots$}	&	\multicolumn{1}{c}{{\bf 3.96$^{4}$}}	&	-15.1	&	$<$6.0	&	4	\\
G334.7225$-$00.6539$^\dagger$	&	16:28:57.91	&	$-$49:36:28.4	&	-44.6	&	-121.1	&	-114.5	&&	3.2	&	12.2	&	5.8	& &	f	&	12.2	&	-138.8	&	5.2	&		\\
G335.7288$-$00.0966$^\dagger$	&	16:30:43.46	&	$-$48:29:39.1	&	-66.2	&	-123.9	&	-133.4	&&	4.4	&	11.1	&	4.9	& &	f	&	11.1	&	-18.8	&	4.3	&		\\
\hline\\
\end{tabular}\\
$^\dagger$ Indicates results obtained from low resolution data.\\
Notes: 1) We indicate the ten sources that are located in a region of Fig.\,\ref{fig:kolpak} that make a distance ambiguous by an ellipsis in the KDA solution column (Col.\,10). 2) Distances that have been assigned using information taken from the literature are discussed in Sect.\,\ref{sect:indivdual_sources} and shown in bold in Col.\,11. 3) If a more reliable distance is available for a particular source or complex we have adopted this value. We identify these sources by adding a superscript to the distances given in Col.\,11; the superscript indicates the reference from which the distance is drawn.  \\
References: (1) \citet{sewilo2004}, (2) \citet{blum2001}, (3) \citet{furness2010} (4) \citet{moises2011}, (5) \citet{xu2011}, (6) \citet{kolpak2003}, (7) \citet{pandian2008}, (8) \citet{watson2003}, (9) \citet{caswell1975}, (10) \citet{green2011}, (11) \citet{busfield2006}, (12) \citet{fish2003}, (13) \citet{caswell1987}, (14) \citet{anderson2009a}, (15) \citet{haynes1979}\\

\end{minipage}

\end{center}
\end{table*}

\setlength{\tabcolsep}{4pt}
\setcounter{table}{2}
\begin{table*}

\begin{center}
\caption{Cont.}

\begin{minipage}{\linewidth}
\scriptsize
\begin{tabular}{lcc...c...cc..cc}

\hline \hline
 & &	  		  		& \multicolumn{3}{c}{Measured \vlsr}& & \multicolumn{3}{c}{Rotation Model} && \multicolumn{4}{c}{Results}   \\ 
\cline{4-6} \cline{8-10}\cline{12-15}
MSX Name&	 	RA  		&Dec & \multicolumn{1}{c}{$v_{\rm{S}}$} &  \multicolumn{1}{c}{$v_{\rm{T}}$} &  \multicolumn{1}{c}{$v_{\rm{A}}$} &&  \multicolumn{1}{c}{Near} & \multicolumn{1}{c}{Far} &  \multicolumn{1}{c}{RGC} && KDA &  \multicolumn{1}{c}{Distance}  & \multicolumn{1}{c}{z} &  \multicolumn{1}{c}{Log(Lum)} & Reference\\
&	(J2000) 	& (J2000) 	&  \multicolumn{1}{c}{(\kms)} & \multicolumn{1}{c}{(\kms)} & \multicolumn{1}{c}{(\kms)} &&   \multicolumn{1}{c}{(kpc)} &  \multicolumn{1}{c}{(kpc)} & \multicolumn{1}{c}{(kpc)} && solution & \multicolumn{1}{c}{(kpc)} & \multicolumn{1}{c}{(pc)} & \multicolumn{1}{c}{(\lsun)}  & \\
\hline

G336.4415$-$00.2597$^\dagger$	&	16:34:21.79	&	$-$48:05:02.7	&	-89.0	&	-123.9	&	-132.8	&&	5.4	&	10.2	&	4.2	& &	f	&	10.2	&	-46.3	&	4.7	&		\\
G336.5396$-$00.1819$^\dagger$	&	16:34:25.10	&	$-$47:57:33.1	&	-84.5	&	-126.8	&	-132.5	&&	5.2	&	10.4	&	4.3	& &	f	&	10.4	&	-33.1	&	4.8	&		\\
G336.8324+00.0301$^\dagger$	&	16:34:40.15	&	$-$47:36:00.3	&	-75.1	&	-126.8	&	-134.6	&&	4.8	&	10.8	&	4.5	& &	f	&	10.8	&	5.7	&	4.8	&		\\
G336.9920$-$00.0244	&	16:35:32.83	&	$-$47:31:09.8	&	-120.9	&	-123.9	&	-128.7	&&	6.8	&	8.8	&	3.5	& &	tp	&	7.8	&	-3.3	&	5.0	&	10	\\
G337.0047+00.3226	&	16:34:05.25	&	$-$47:16:30.7	&	-62.8	&	-126.8	&	-122.2	&&	4.3	&	11.4	&	4.9	& &	f	&	11.4	&	64.2	&	4.9	&		\\
G337.1218$-$00.1748	&	16:36:43.41	&	$-$47:31:28.5	&	-75.1	&	-123.9	&	-128.7	&&	4.8	&	10.9	&	4.5	& &	f	&	10.9	&	-33.1	&	5.8	&	13	\\
G337.6651$-$00.1750	&	16:38:52.22	&	$-$47:07:16.3	&	-53.1	&	-126.8	&	-132.0	&&	3.8	&	11.9	&	5.2	& &	f	&	11.9	&	-36.4	&	5.0	&		\\
G337.7051$-$00.0575	&	16:38:30.79	&	$-$47:00:46.4	&	-48.3	&	-126.8	&	-122.2	&&	3.6	&	12.2	&	5.4	& &	f	&	12.2	&	-12.2	&	$<$5.4	&	11,12	\\
G337.7091+00.0932	&	16:37:52.29	&	$-$46:54:33.1	&	-76.7	&	-126.8	&	-127.4	&&	4.9	&	10.8	&	4.4	& &	f	&	10.8	&	17.6	&	4.5	&	10	\\
G337.9266$-$00.4588$^\dagger$	&	16:41:08.30	&	$-$47:06:50.7	&	-40.2	&	-129.6	&	-124.6	&&	3.1	&	12.6	&	5.7	& &	f	&	12.6	&	-101.0	&	6.3	&	13	\\
G338.3340+00.1315	&	16:40:07.96	&	$-$46:25:04.0	&	-37.3	&	-129.6	&	-106.4	&&	3.0	&	12.8	&	5.8	& &	f	&	12.8	&	29.4	&	5.4	&		\\
G338.4033+00.0338$^\dagger$	&	16:40:49.44	&	$-$46:25:50.5	&	-40.3	&	-129.6	&	-129.6	&&	3.2	&	12.6	&	5.7	& &	f	&	12.6	&	7.5	&	4.6	&		\\
G338.4357+00.0591	&	16:40:50.32	&	$-$46:23:22.9	&	-33.5	&	-129.6	&	-129.4	&&	2.8	&	13.1	&	6.0	& &	f	&	13.1	&	13.5	&	$<$6.1	&		\\
G338.9173+00.3824	&	16:41:16.65	&	$-$45:48:52.9	&	-24.7	&	-129.6	&	-76.9	&&	2.2	&	13.7	&	6.5	& &	\multicolumn{1}{c}{$\cdots$}	&	\multicolumn{1}{c}{$\cdots$}	&	\multicolumn{1}{c}{$\cdots$}	&	\multicolumn{1}{c}{$\cdots$}	&		\\
G338.9217+00.6233	&	16:40:15.62	&	$-$45:39:06.4	&	-62.1	&	-129.6	&	-73.6	&&	4.3	&	11.5	&	4.7	& &	n	&	4.3	&	47.1	&	4.5	&	13	\\
G338.9341$-$00.0623$^\dagger$	&	16:43:16.03	&	$-$46:05:42.0	&	-43.9	&	-132.5	&	-50.6	&&	3.4	&	12.5	&	5.4	& &	n	&	3.4	&	-3.7	&	3.6	&	10	\\
G339.1052+00.1490	&	16:42:59.81	&	$-$45:49:37.9	&	-78.2	&	-132.5	&	-83.5	&&	5.0	&	10.9	&	4.2	& &	n	&	5.0	&	13.1	&	4.3	&		\\
G339.5836$-$00.1265$^\dagger$	&	16:45:59.04	&	$-$45:38:42.7	&	-34.1	&	-135.4	&	-136.2	&&	2.9	&	13.1	&	5.9	& &	f	&	13.1	&	-28.8	&	4.5	&	10, 13	\\
G340.2480$-$00.3725	&	16:49:30.14	&	$-$45:17:48.4	&	-50.3	&	-135.4	&	-53.9	&&	3.9	&	12.1	&	5.0	& &	n	&	3.9	&	-25.1	&	4.5	&	10	\\
G340.2490$-$00.0460	&	16:48:05.25	&	$-$45:05:09.6	&	-122.0	&	-135.4	&	-132.7	&&	6.7	&	9.3	&	3.1	& &	f?	&	9.3	&	-7.4	&	5.1	&	10	\\
G340.2768$-$00.2104$^\dagger$	&	16:48:54.11	&	$-$45:10:14.1	&	-45.5	&	-135.4	&	-95.6	&&	3.6	&	12.4	&	5.2	& &	\multicolumn{1}{c}{$\cdots$}	&	\multicolumn{1}{c}{$\cdots$}	&	\multicolumn{1}{c}{$\cdots$}	&	\multicolumn{1}{c}{$\cdots$}	&		\\
G342.0610+00.4200$^\dagger$	&	16:52:33.57	&	$-$43:23:42.3	&	-65.0	&	-141.3	&	-113.2	&&	3.5	&	12.7	&	5.3	& &	\multicolumn{1}{c}{$\cdots$}	&	\multicolumn{1}{c}{$\cdots$}	&	\multicolumn{1}{c}{$\cdots$}	&	\multicolumn{1}{c}{$\cdots$}	&		\\
G343.5024$-$00.0145$^\dagger$	&	16:59:20.90	&	$-$42:32:38.4	&	-27.7	&	-147.2	&	-123.0	&&	2.8	&	13.5	&	5.9	& &	f	&	13.5	&	-3.4	&	5.6	&		\\
G344.2207$-$00.5953	&	17:04:13.32	&	$-$42:19:57.3	&	-23.2	&	-147.2	&	-34.9	&&	2.5	&	13.9	&	6.1	& &	n	&	2.5	&	-26.0	&	4.6	&	10	\\
G344.3976+00.0533$^\dagger$	&	17:02:02.04	&	$-$41:47:48.1	&	-65.9	&	-147.2	&	-85.7	&&	5.0	&	11.4	&	4.0	& &	n	&	5.0	&	4.6	&	3.9	&		\\
G344.4257+00.0451	&	17:02:09.65	&	$-$41:46:46.2	&	-66.3	&	-147.2	&	-87.4	&&	5.0	&	11.4	&	3.9	& &	n	&	5.0	&	3.9	&	4.9	&		\\
G345.4881+00.3148	&	17:04:28.17	&	$-$40:46:22.4	&	-17.6	&	-150.2	&	-31.0	&&	2.1	&	14.3	&	6.5	& &	n	&	2.1	&	11.7	&	4.9	&	1	\\
G345.5285$-$00.0508$^\dagger$	&	17:06:08.34	&	$-$40:57:43.2	&	-3.7	&	-153.2	&	-125.3	&&	0.5	&	16.0	&	8.1	& &	\multicolumn{1}{c}{$\cdots$}	&	\multicolumn{1}{c}{{\bf 16.0}}	&	-14.2	&	5.2	&	13	\\
G345.5472$-$00.0801	&	17:06:19.34	&	$-$40:57:52.9	&	-6.0	&	-150.2	&	-118.9	&&	0.8	&	15.7	&	7.8	& &	\multicolumn{1}{c}{$\cdots$}	&	\multicolumn{1}{c}{{\bf 15.7}}	&	-21.9	&	4.7	&	13	\\
G345.6495+00.0084	&	17:06:16.48	&	$-$40:49:46.9	&	-10.8	&	-150.2	&	-132.7	&&	1.4	&	15.1	&	7.2	& &	f	&	15.1	&	2.2	&	5.7	&	13	\\
G347.5998+00.2442	&	17:11:21.91	&	$-$39:07:27.1	&	-94.8	&	-156.2	&	-96.6	&&	6.3	&	10.4	&	2.7	& &	n	&	6.3	&	26.7	&	4.3	&	10	\\
G348.5312$-$00.9714	&	17:19:15.28	&	$-$39:04:31.0	&	-14.1	&	-159.3	&	-34.3	&&	2.1	&	14.6	&	6.4	& &	n	&	\multicolumn{1}{c}{2.84$^{4}$}	&	-48.1	&	4.2	&	4, 13	\\
G348.6972$-$01.0263	&	17:19:58.55	&	$-$38:58:14.5	&	-12.8	&	-159.3	&	-36.2	&&	2.0	&	14.7	&	6.6	& &	n	&	\multicolumn{1}{c}{2.84$^{4}$}	&	-50.9	&	4.2	&	4, 13	\\
G348.7121+00.3279$^\dagger$	&	17:14:22.03	&	$-$38:10:31.0	&	-6.6	&	-162.3	&	-172.9	&&	1.1	&	15.6	&	7.4	& &	f	&	15.6	&	89.2	&	\multicolumn{1}{c}{$\cdots$}	&		\\
G348.7250$-$01.0435	&	17:20:07.82	&	$-$38:57:27.7	&	-12.7	&	-159.3	&	-34.3	&&	2.0	&	14.7	&	6.6	& &	n	&	\multicolumn{1}{c}{2.84$^{4}$}	&	-51.7	&	$<$4.7	&	4, 13	\\
\hline\\
\end{tabular}\\
$^\dagger$ Indicates results obtained from low resolution data.\\
Notes: 1) We indicate the ten sources that are located in a region of Fig.\,\ref{fig:kolpak} that make a distance ambiguous by an ellipsis in the KDA solution column (Col.\,10). 2) Distances that have been assigned using information taken from the literature are discussed in Sect.\,\ref{sect:indivdual_sources} and shown in bold in Col.\,11. 3) If a more reliable distance is available for a particular source or complex we have adopted this value. We identify these sources by adding a superscript to the distances given in Col.\,11; the superscript indicates the reference from which the distance is drawn.  \\
References: (1) \citet{sewilo2004}, (2) \citet{blum2001}, (3) \citet{furness2010} (4) \citet{moises2011}, (5) \citet{xu2011}, (6) \citet{kolpak2003}, (7) \citet{pandian2008}, (8) \citet{watson2003}, (9) \citet{caswell1975}, (10) \citet{green2011}, (11) \citet{busfield2006}, (12) \citet{fish2003}, (13) \citet{caswell1987}, (14) \citet{anderson2009a}, (15) \citet{haynes1979}\\

\end{minipage}

\end{center}
\end{table*}
\setlength{\tabcolsep}{6pt}

\subsection{Notes on specific sources}
\label{sect:indivdual_sources}

In the previous subsection we identified ten sources that are located
in a region of the plot presented in Fig.\,5 that makes allocating a
distance problematic. We have conducted a literature and SIMBAD search
to find complementary information and/or associations with known giant
molecular cloud (GMC) complexes to help break the distance ambiguities
towards these sources. As a result of this analysis we were able to
associate seven sources with one of four well known complexes. In this subsection we
present a summary of this review and the assigned distance.

\subsubsection{W31-South Complex: G010.1615$-$00.3623}

The \HI\ spectrum shows clear evidence of distinct absorption between the tangent and source velocities; this
would suggest this source is located at the far distance, however, the absorption stops far short of the velocity of the tangent which makes a distance determination problematic. \citet{sewilo2004} observed H110$\alpha$ and H$_2$CO towards this source using the NRAO Green Bank Telescope and obtained similar velocities. They placed this source at
the far distance but noted that the non-detection of \HI\ absorption at the tangent point velocity presented a problem. Fortunately, there have have been a number of spectroscopic studies that have determined distances to W31; both \citet{blum2001} and \citet{furness2010} determined similar distances of 3.4\,kpc and $\sim$3.3\,kpc, respectively. More recently \citet{moises2011} determined the distances to W31-South and W31-North to be 3.55 and 2.39\,kpc using spectrophotometric measurements. These distances are in reasonable agreement with the near kinematic solution of 2.5$\pm1.6$\,kpc. 

The position and velocity of G010.1615$-$00.3623 would place it in W31-South complex of molecular clouds and bright \HII~regions and we have therefore allocated a distance of 3.55\,kpc to this source.

\subsubsection{RCW\,106 Complex}

Comparing the angular projection and velocities of our sample we find 10 sources are embedded within this giant molecular cloud associated with the RCW\,106 star forming region ($l\sim$333\degr, $b\sim$$-$0.5\degr). \citet{bains2006} mapped the molecular structure of the whole complex using $^{13}$CO. Examination of these data reveals all of these sources are associated with this complex (i.e., they are all connected in $lbv$ space). This sample includes two sources we have so far been unable to allocate a distance for; these are G332.5438$-$00.1277 and G332.6450$-$00.6036. Of the remaining eight sources we have placed six at the near distance and two at the far distance (i.e., G332.1544-00.4487 and G333.6032$-$00.2184). The far distance allocation was determined using the lower resolution \HI\ data which is less reliable (see discussion Sect.\,\ref{sect:hi_low_comparison} for more detail) than the high resolution data and therefore it is more likely that this complex is located at the near distance. 

The small differences in radial velocity across the complex (of order a few \kms) results in a range of near kinematic distances of 3.3-3.8\,kpc, which is consistent with the distance of 3.6\,kpc determined by \citet{lockman1979} and the spectrophotometric distance of 3.96\,kpc determined by \citet{moises2011}. It is the latter of these distances we have allocated to all of the sources associated with this complex.

\subsubsection{G332.783+0.792}

G333.0058+00.7707 is positionally associated with the \HII~region G332.783+0.792 observed by \citet{caswell1987} who placed this source at the near distance. The source velocity determined from CO observations is $-48.3$\,\kms, which compares very well to the velocity of the H$_2$CO absorption feature at $-48.4$\,\kms and the RRL velocity of $-52$\,\kms\ reported by \citet{caswell1987}. It is likely that G333.0058+00.7707 is physically associated with this \HII~region and we have therefore adopted the near distance allocated by \citet{caswell1987} for this source.

\subsubsection{G345.555$-$0.042 Complex}

G345.5285$-$00.0508 and G345.5472$-$00.0801 are located in the G345.555$-$0.042 GMC complex. The \HII~region associated with this complex was observed by \citet{caswell1987} and found to have a RRL velocity of $-$6\,\kms. The highest velocity H$_2$CO component was measured at $-$15\,\kms, which led them to place this source at the far distance. We also adopt this kinematic distance for this complex.

\subsection{Comparison of high and low resolution data}
\label{sect:hi_low_comparison}

\begin{figure}
\includegraphics[width=0.95\linewidth, trim= 0 0 0 0]{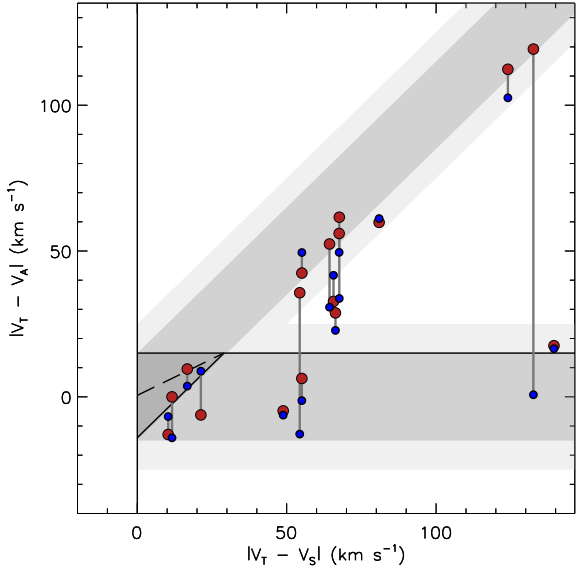}

\caption{\label{fig:kolpak_comp} Plot of the velocity differences of the nineteen \HII~regions for which both high and low resolution data is available; the high and low resolution data is shown in red and blue respectively, with matching sources being connected by a straight line.} 
\end{figure}

High- and low-resolution data are available for seventeen sources and it is therefore possible to compare the results obtained from different angular resolution data for consistency. In Fig.\,\ref{fig:kolpak_comp} we present a plot showing the velocity differences obtained for the overlapping high- and low-resolution samples (coloured red and blue respectively); the  high- and low-resolution data for each source are linked by a solid line. The source and tangent velocities are independent of the resolution of the \HI\ data for any given source and so the x-axis position for each source is unchanged. The choice of angular resolution only affects the measured velocity of the absorption feature.

In Table\,\ref{tbl:KDA_diff} we provide a summary of the kinematic distance solutions derived from the high and low resolution data sets as well as the source and tangent velocities, and the velocity offset between the two measurements of the absorption features. Comparing the results for the high and low resolution data reveals them to be in reasonable agreement with a velocity difference between the absorption features measured from the high and low resolution data ($|\Delta v_{\rm{A}}|$) less than 10\,\kms\ in 11 cases ($\sim$60\,per\,cent). Looking at the kinematic distance solution derived from the different resolution data sets we find agreement in 11 of the 17 sources in common.

\setlength{\tabcolsep}{3pt}
\begin{table}



\begin{center}
\caption{Comparison of kinematic distance solutions derived from the high and low resolution data for the 17 sources common to both data sets.}
\label{tbl:KDA_diff}
\begin{minipage}{\linewidth}
\begin{tabular}{lcc....}

\hline \hline
 & & &&\multicolumn{3}{c}{Measured Velocities}\\ 
 \cline{5-7}

MSX Name&	 	\multicolumn{2}{c}{KDA Solution}  		&& \multicolumn{1}{c}{$v_{\rm{S}}$} &  \multicolumn{1}{c}{$v_{\rm{T}}$} &  \multicolumn{1}{c}{$\Delta v_{\rm{A}}$} \\
\cline{2-3}
&	 ATCA &SGPS 	& &  \multicolumn{1}{c}{(\kms)} & \multicolumn{1}{c}{(\kms)} & \multicolumn{1}{c}{(\kms)}\\
\hline
G307.5606$-$00.5871	&	f	&	f	&&	-33.0	&	-54.3	&	-15.02	\\
G333.3072$-$00.3666$^\dagger$	&	?	&	n?	&&	-49.9	&	-115.5	&	-8.97	\\
G332.8256$-$00.5498	&	n	&	n	&&	-57.7	&	-112.8	&	-7.04	\\
G328.3067+00.4308	&	f	&	f	&&	-91.7	&	-102.0	&	-6.16	\\
G344.4257+00.0451	&	n	&	n	&&	-66.3	&	-147.2	&	-1.30	\\
G345.6495+00.0084	&	f	&	f	&&	-10.8	&	-150.2	&	1.02	\\
G337.1218$-$00.1748	&	f	&	f	&&	-75.1	&	-123.9	&	1.46	\\
G307.6213$-$00.2622	&	f?	&	f	&&	-37.6	&	-54.3	&	5.85	\\
G333.6032$-$00.2184$^\dagger$	&	?	&	f?	&&	-49.2	&	-115.5	&	6.01	\\
G338.9217+00.6233	&	n	&	n	&&	-62.1	&	-129.6	&	6.48	\\
G318.9148$-$00.1647	&	f	&	f	&&	-22.2	&	-77.3	&	7.60	\\
G344.2207$-$00.5953	&	n	&	n	&&	-23.2	&	-147.2	&	9.79	\\
G305.3500+00.2240	&	f?	&	f	&&	-39.0	&	-50.6	&	14.04	\\
G333.1306$-$00.4275$^\dagger$	&	n	&	?	&&	-51.2	&	-115.5	&	21.67	\\
G333.0162+00.7615$^\dagger$	&	n	&	?	&&	-47.9	&	-115.5	&	27.86	\\
G326.7249+00.6159$^\dagger$	&	n	&	f	&&	-42.5	&	-96.8	&	48.44	\\
G345.4881+00.3148$^\dagger$	&	n	&	f	&&	-17.6	&	-150.2	&	118.53	\\

\hline\\
\end{tabular}\\
$^\dagger$ Indicates sources where the distance determined separately from the high and low resolution data disagree.\\

\end{minipage}

\end{center}
\end{table}
\setlength{\tabcolsep}{6pt}

\begin{figure*}
\begin{center}
\includegraphics[width=0.49\textwidth, trim= 0 0 0 0]{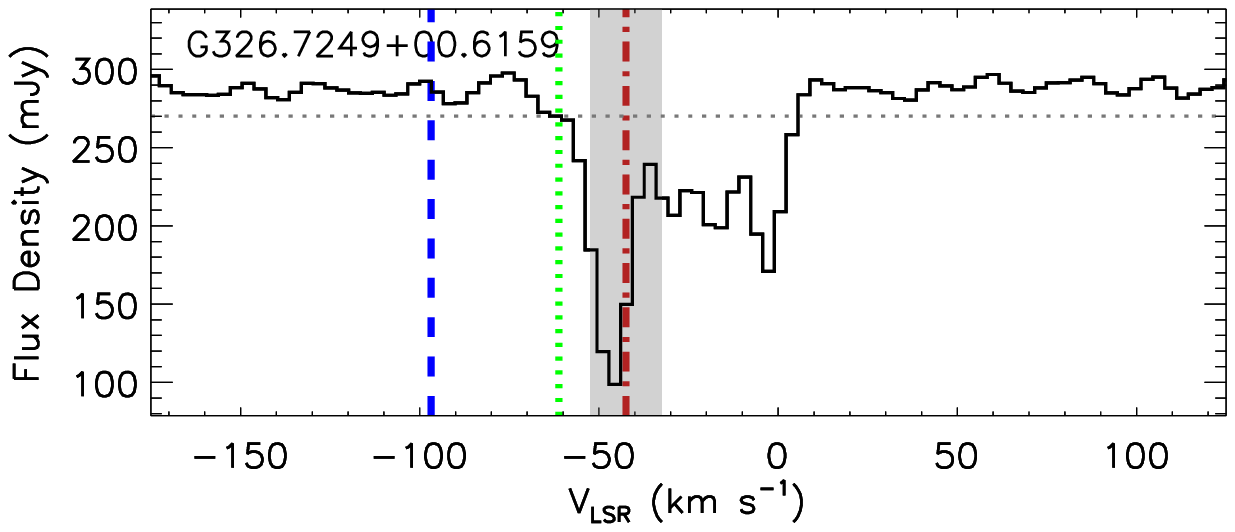}
\includegraphics[width=0.49\textwidth, trim= 0 0 0 0]{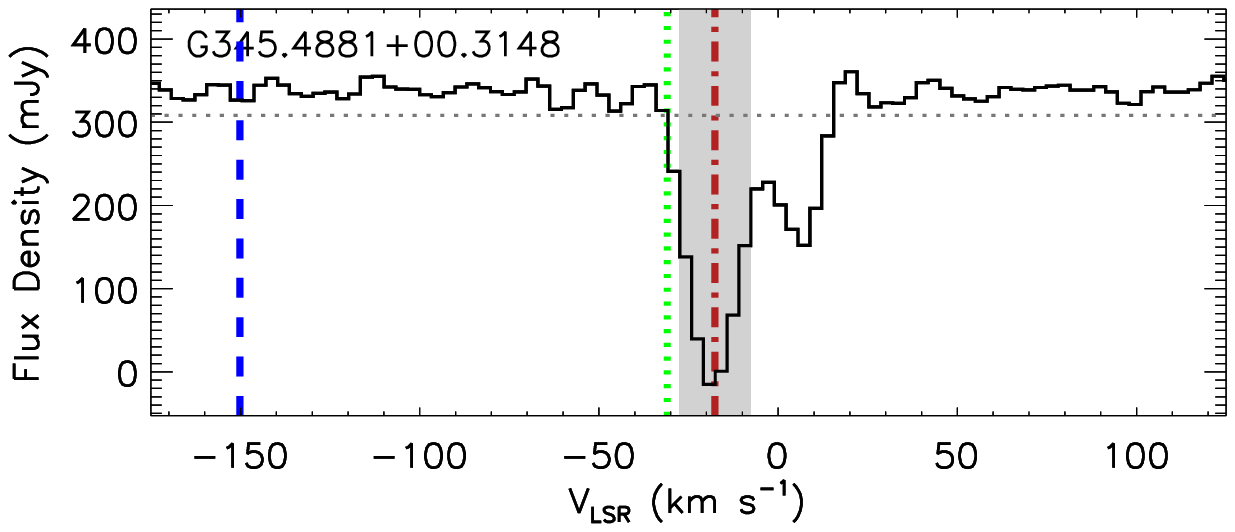}
\includegraphics[width=0.49\textwidth, trim= 0 0 0 0]{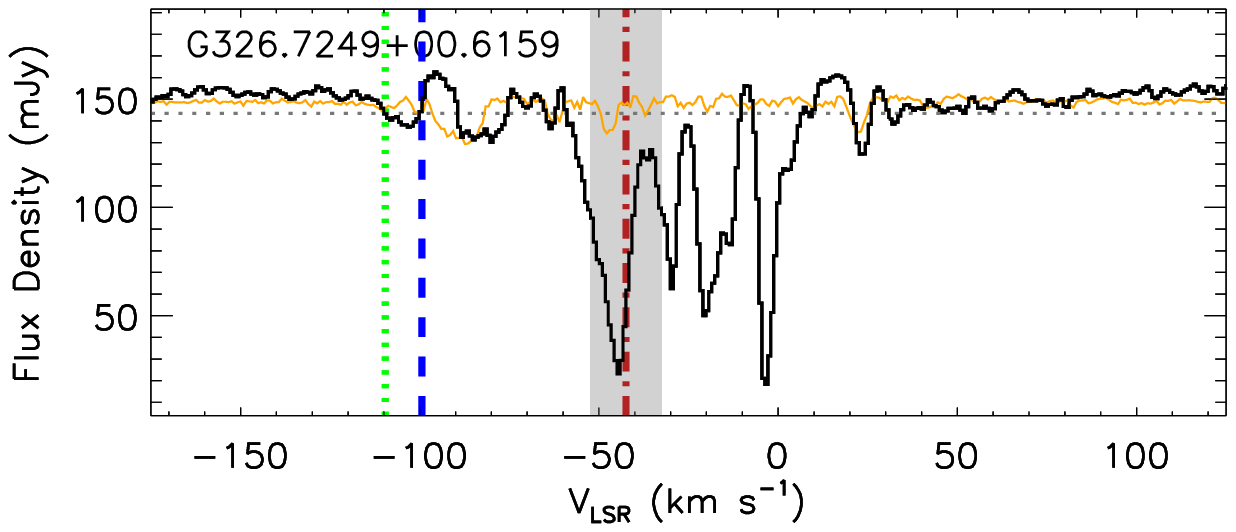}
\includegraphics[width=0.49\textwidth, trim= 0 0 0 0]{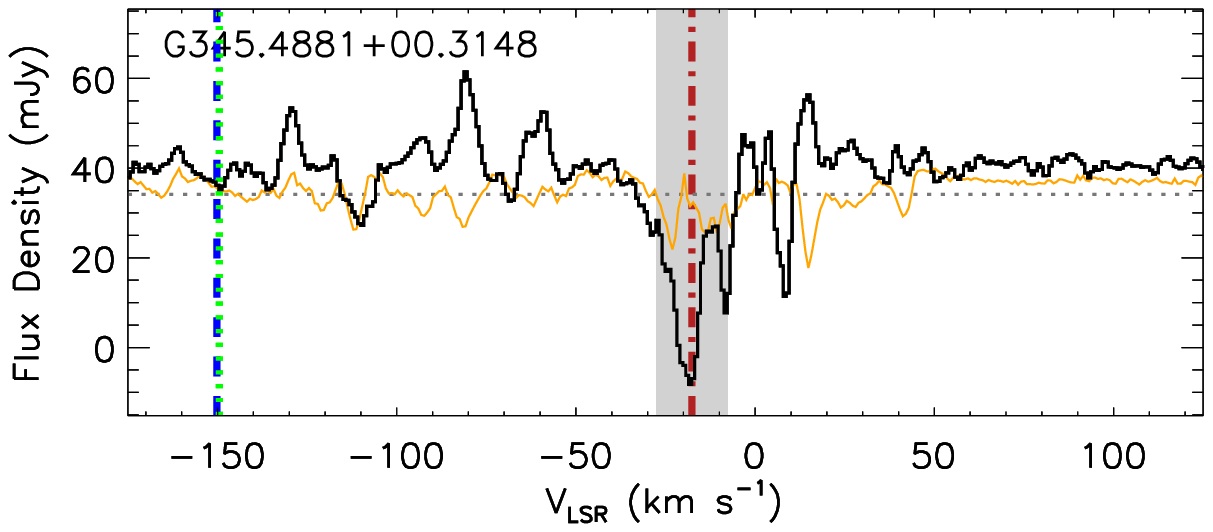}

\caption{\label{fig:KDA_diff} \HI\ spectra taken towards the two source where distance determined from the high and low spatial resolution data disagree. In the top and bottom panels we present the ATCA and SGPS spectra, respectively. The source velocity ($v_{\rm{s}}$), the velocity of the tangent point ($v_{\rm{t}}$) and the position of the first absorption minimum ($v_{\rm{a}}$) are shown by the red, blue and green vertical lines, respectively. The grey vertical band covers the velocity region 10\,\kms\ either side of the source velocity and is provided to give an indication of the uncertainty associated with it due to streaming motions. The dotted horizontal and solid yellow lines shows the 4$\sigma_{\rm{r.m.s}}$ receiver noise level determined from absorption free parts of the spectra and the \HI\ emission fluctuations, respectively.}

\end{center}
\end{figure*}

We identify the six sources where the distance allocation differs by appending a $\dagger$ to the MSX name in Table\,\ref{tbl:KDA_diff}. For four of these we are only able to resolve the distance ambiguity using one of the available data sets. This leaves only two sources (i.e., G326.7249+00.6159  and G345.4881+00.3148) where the differences in the velocity of the absorption features seen in the spectra have resulted in incorrect distances being assigned. In Fig.\,\ref{fig:KDA_diff} we present the high and low resolution \HI\ spectra for these two sources.

For both sources the structure of the absorption features seen in the high-resolution data are broadly repeated in the lower resolution SGPS spectra, however, the lower-resolution spectra also show a number of additional absorption dips lying between the source velocity and the tangent velocity. The higher signal-to-noise ratio of the ATCA data and the fact that it is relatively unaffected by contamination from diffuse continuum emission make the near-distance solution for these two sources very reliable. We therefore conclude that the additional absorption features seen in the SGPS spectrum are not associated with either source. The cause of these additional absorption features seen in the SGPS data is unclear but it is possibly they are the result of fluctuations in the background and/or noise. 

We notice that, for five of the six sources where the distance allocation determined from the high- and low-resolution data, the lower-resolution data have resulted in absorption features being detected significantly closer to the the tangent velocity than found for the high-resolution spectra. We would therefore conclude that these anomalous absorption features associated with the SGPS data are quite common and as a consequence the distance allocations made using these data are less reliable than found using higher-resolution data. It is hard to estimate the impact this uncertainty may have on the overall reliability due to the small sample of sources for which we have both high- and lower-resolution data available. However, this could explain the $\sim$20\,per\,cent disagreement often found from similar comparisons reported in the literature (e.g., Green et al. 2011, \citealt{anderson2009a}).

\subsection{Comparison with previous studies}
\label{sect:lit_comparison}

Resolving distance ambiguities is crucial for many aspects of Galactic astronomy but particularly for understanding large-scale structure of the Milky Way. As a consequence solving these ambiguities has become an intense area of research in recent years which has resulted in a number of publications. To check the reliability of our distance assignments we have compared our results with those previously reported in the literature.

\setlength{\tabcolsep}{4pt}
\begin{table}


\begin{center}
\caption{Distance disagreements.}
\label{tbl:distance_disagreements}
\begin{minipage}{\linewidth}
\begin{center}
\begin{tabular}{lccc}

\hline \hline
MSX Name&	 	\multicolumn{1}{c}{This Paper}  		& \multicolumn{1}{c}{Literature} &  \multicolumn{1}{c}{Reference} \\
\hline
G320.1750+00.8001$^\dagger$	&	f	&	n	&	1\\
G321.0523$-$00.5070$^\dagger$	&	f	&	n	&	1\\
G328.5739$-$00.5483$^\dagger$	&	f	&	n	&	1\\
G330.8708$-$00.3715$^\dagger$	&	f	&	n	&	1\\
G330.9544$-$00.1817	&	f?	&	n	&	2\\
G331.5414$-$00.0675$^\dagger$	&	f	&	n	&	1\\
G337.7091+00.0932	&	f	&	n	&	2\\
G337.9266$-$00.4588$^\dagger$	&	f	&	n	&	1\\
G339.5836$-$00.1265$^\dagger$	&	f	&	n,f,f	&	1, 2, 3\\
G340.2480$-$00.3725	&	n	&	f	&	2\\
\hline\\

\end{tabular}\\
\end{center}
$^\dagger$ Indicates results obtained from low resolution data.\\
References: (1) \citet{caswell1987}, (2) \citet{green2011}, (3) \citet{haynes1979}\\

\end{minipage}

\end{center}
\end{table}
\setlength{\tabcolsep}{6pt}

Excluding a single source placed at the tangent point we find 51 of our sample have distances previously reported in the literature (see last column in Table\,\ref{tbl:analysis_results} for references). Of these we find agreement between the distances derived in this paper and those given in the literature in 41 cases, which corresponds to $\sim$80\,per\,cent of the sample. However, there are a significant number of cases where our distance allocations disagree with those in the literature; in Table\,\ref{tbl:distance_disagreements} we present a summary of these sources along with the distance determined in this paper and the distance previously determined in the literature. 

We find that distances of 7 of the 10 sources presented in Table\,\ref{tbl:distance_disagreements} have been determined using the SGPS data. Moreover, in all of these cases we find that the SGPS data suggests a far distance allocation therefore the anomalous absorption features discussed in the previous subsection could be a large contributing factor. We share these sources with three other studies; three with a study based on \HI\ self-absorption using methanol masers to trace the velocity of the star formation regions (\citealt{green2011}), six sources with a radio recombination line (RRL) emission survey of southern \HII~regions (\citealt{caswell1987}), and one source (G339.5836$-$00.1265) which is included in both of these and \citet{haynes1979}.

\citet{green2011} has used the SGPS continuum subtracted data to look for \HI\ self-absorption at a similar velocity to the methanol maser velocity to resolve the distance ambiguity to a large sample of sources. Of the four sources we share with \citet{green2011} three have been observed at high resolution ($\sim$10\arcsec) with the ATCA and we therefore consider the distances allocated using this data more reliable. Indeed \citet{green2011} place a lower confidence for all of the sources that are shared between the two surveys. The fourth source  (G339.5836$-$00.1265) we share with the sample of \citet{green2011} (G339.582$-$0.127) has also been observed by \citet{caswell1987} and at higher resolution by \citet{haynes1979} (G339.578$-$0.124). The velocity of the peak methanol maser, CO emission and RRL are $-$30.4, $-$34.1 and $-$30\,\kms, respectively, and thus are all likely to be associated with the same region. Both \citet{caswell1987} and \citet{haynes1979} place this source at the far distance, which agrees with our evaluation, however, Green et al. places it at the near distance. Given the data available we consider the far distance to be more likely.

In total we share 21 sources in common with  \citet{caswell1987} and disagree with their distance allocation in six cases ($\sim$28\,per\,cent). The main criterion used by \citet{caswell1987} to determine if a particular source was located at the near distance was by association with an optical counterpart. This was based on the fact that it is not generally possible to detect the optical counterparts for sources more distant than $\sim$6\,kpc. However, \citet{caswell1987} note that this may sometimes be wrong if there is a chance alignment of a nearby optical nebula with a more distant \HII~region. 

This proportion of disagreement with \citet{caswell1987} is similar to that reported by \citet{green2011}  and \citet{anderson2009a}. The agreement between this work and previous studies is extremely good and we believe that the limits are a fair representation of the inherent uncertainly associated with the  method we have used to resolve distance ambiguities.

\section{Galactic structure}
\label{sect:discussion}

We have been able to assign a distance to 102 of the 105 \HII~regions
in our sample, with 39 (33 with high confidence) sources being placed
at the near distance and 60 (49 with high confidence) being placed at
the far distance. Three \HII~regions were located at the distance of the
tangent point and we were unable to resolve the distance ambiguity for four sources. 

In this section we will use the distance and luminosity results to investigate the spatial distribution of this sample of \HII~regions with respect to large-scale structure features of the Milky Way. Optical observations of spiral galaxies reveal an intimate relationship between spiral arms and the galactic population of young massive stars and their associated \HII~regions. Regions of massive star formation are almost exclusively found to be associated with the spiral arms (\citealt{kennicutt2005}) where molecular clouds are thought to form at the leading edges of spiral arms from gas compressed through spiral density waves (\citealt{roberts1969}). The Galactic distribution of massive young stars could, therefore, be an important probe of Galactic structure. 

In a separate paper we have estimated the bolometric fluxes of a large number of young massive stars identified by the RMS survey. This has been done by fitting stellar models to each source's spectral energy distribution using infrared to millimetre flux measurements (see \citealt{mottram2010,mottram2011a} for details). Combining these bolometric flux values with the distances determined in the previous section allows us to estimate the bolometric luminosity of each \HII~region. The calculated bolometric luminosities can be found in the Col.\,13 of Table\,\ref{tbl:analysis_results}. 

\begin{figure*}
\includegraphics[width=0.99\linewidth, trim= 0 0 0 0]{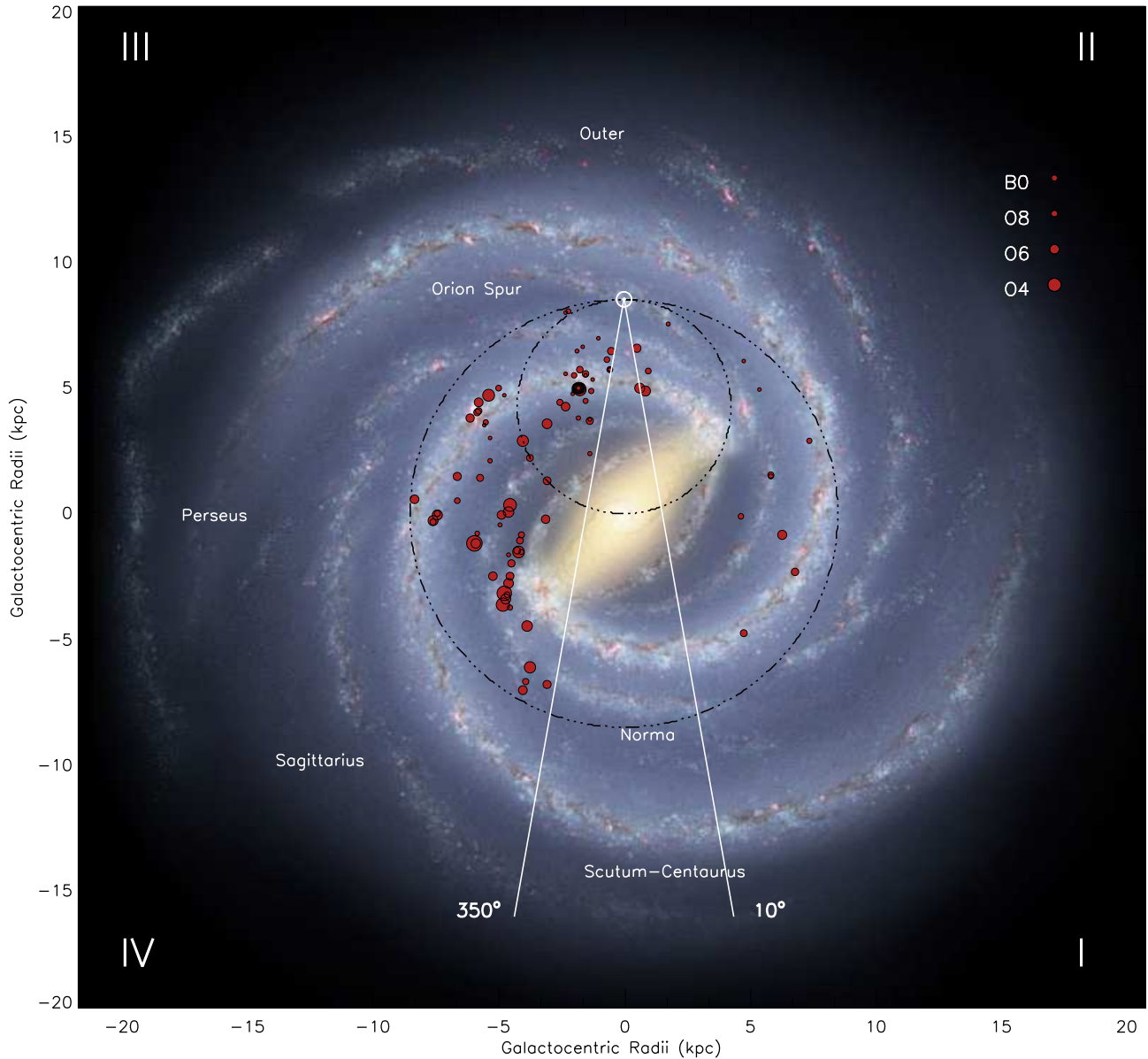}

\caption{\label{fig:gal_dist} Galactic distribution of the RMS selected sample of UC \HII~regions located within the solar circle presented in this paper. The distribution of \HII~regions is superimposed over a sketch of how the Galaxy is thought to appear if viewed face-on from the north Galactic pole with Galactic centre at (0,0)\,kpc and the Sun at (0,8.5)\,kpc; the positions of the Galactic centre and Sun are indicated by the small circle. This image has been produced by Robert Hurt of the Spitzer Science Center in consultation with Robert Benjamin and attempts to synthesise many of the key elements of Galactic structure using the best data currently available (see text for more details). The positions of the \HII~regions are shown as red-in-black circles the sizes of these circles give an indication of their respective luminosities. In the upper right corner we give the luminosities for a sample of zero age main sequence stars. The Roman numerals in the corners refer to the Galactic quadrants and the two thick white lines originating from the location of the Sun enclose the region of the Galactic Plane excluded from the RMS survey ($350\degr < l < 10\degr$).  The dot-dashed circles represent the locus of tangent points and the Solar Circle.}

\end{figure*}

In an effort to examine the distribution of this sample of \HII~regions with respect to the large-scale structure of the  Milky Way, we plot their projected positions in Fig.\,\ref{fig:gal_dist} over an image of the Galaxy. The size of the symbols is proportional to each source's bolometric luminosity and, for \HII~regions that are associated with a complex with a known distance, the complex distance has been adopted. The background image used in this figure has been produced by Robert Hurt of the Spitzer Science Center in consultation with Robert Benjamin (University of Wisconsin-Whitewater) and attempts to synthesise all that has been learnt about Galactic structure over the past fifty years including: a 3.1-3.5\,kpc Galactic Bar at an angle of 20\degr\ with respect to the Galactic Centre-sun axis (\citealt{binney1991,blitz1991,dwek1995}), a second non-axisymmetric structure referred to as the ``Long Bar'' (\citealt{hammersley2000}) with a Galactic radius of $4.4\pm0.5$\,kpc at an angle of 44\degr\ $\pm 10$\degr\ \citep{benjamin2005}, the Near and Far 3-kpc arms, and the four principle arms: Norma, Sagittarius, Perseus and Scutum-Centaurus. The position of the arms is based on the \citet{georgelin1976} model which has been modified to incorporate Very Long Baseline Array maser parallax measurements (e.g., \citealt{xu2006}) and refined directions for the spiral arm tangents from \citet{dame2001}. The Perseus and Scutum-Centaurus arms have been emphasised in this image to reflect the overdensities seen in the old stellar disk population towards their expected Galactic longitudes tangent positions \citep{benjamin2008,churchwell2009}.  

The number of \HII~regions with resolved distance ambiguities presented
here is not sufficient to expect to see evidence of ``well-defined''
spiral structure without something to guide the eye. However, it is
clear from a visual examination of Fig.\,\ref{fig:gal_dist} that the
\HII~regions are tightly correlated with the expected position of the
spirals, with a large number of \HII~region lying on or near an arm
and the inter-arm regions much more sparsely populated.

\begin{figure}
\includegraphics[width=0.99\linewidth, trim= 0 0 0 0]{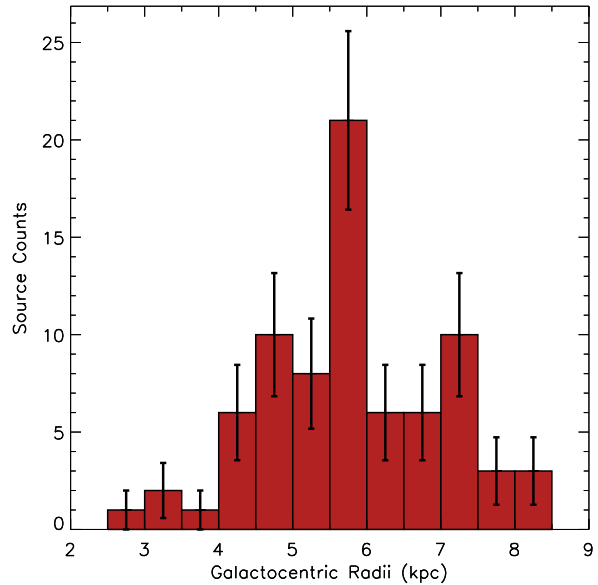}

\caption{\label{fig:gal_rgc_dist} Plot showing the distribution of \HII~regions as a function of Galactocentric radius. Errors have been calculated assuming \poi\ statistics (i.e., $\sqrt{N}$ where $N$ is the number of sources in each bin).}

\end{figure}

In Fig.\,\ref{fig:gal_rgc_dist} we show the distribution of the southern sample of \HII~regions as a function of Galactocentric radius. We have exclude the northern \HII~regions as there are too few to draw any reliable conclusions and their distribution has already been discussed by \cite{urquhart2011}. To avoid any Malmquist-type bias we only include  \HII~regions with bolometric luminosities above our completeness limit ($\sim$10$^4$\,\lsun; \citealt{mottram2011a}). The distribution of the southern sample of \HII~regions shown in Fig.\,\ref{fig:gal_rgc_dist} reveals the presence of a strong peak at a Galactocentric radius between approximately 5.5-6\,kpc. A significant contribution to this peak in source counts is provided by a cluster of luminous \HII~regions that are positionally coincident with the southern end of the Galactic Long Bar, which is located at a Galactic longitude of $\sim$340\degr. A similar increase in the RMS source density is seen in the northern Galactic plane at a Galactocentric radius of $\sim$4\,kpc, which is positionally coincident with the intersection of the Long Bar and the Scutum-Centaurus arm (see \citealt{urquhart2011} for details). There is also some evidence of a second peak in the radial distribution between  7-7.5\,kpc, which is roughly coincident with the Galactocentric radius of the Scutum-Centaurus arm, however, a larger sample will be required before we can determine whether this peak is significant. 

The overall structure of the Galactocentric distribution found for the northern and southern Galactic plane is markedly different from each other (cf. Fig.\,10 of \citealt{urquhart2011}). Our analysis of the Galactocentric distribution of the RMS sample of \HII~regions and MYSOs located in the northern Galactic plane identified three strong peaks at approximately 4, 6 and 8\,kpc; these peaks were shown to be coincident with the intersection of the Long Bar and the Scutum-Centaurus arm, and Galactocentric radii of the Sagittarius and Perseus arms, respectively. However, the detection of a single strong peak in the Galactocentric distribution of the southern Galactic plane would suggest that the overall structure is different. The distribution with Galactocentric radius is only dependent on the Galactic rotation curve and not on the solution of near-far ambiguity, and therefore, the differences in Galactocentric radii of peaks seen in the northern and southern Galactic planes are significant.

\begin{figure}
\begin{center}
\includegraphics[width=0.99\linewidth, trim= 0 0 0 0]{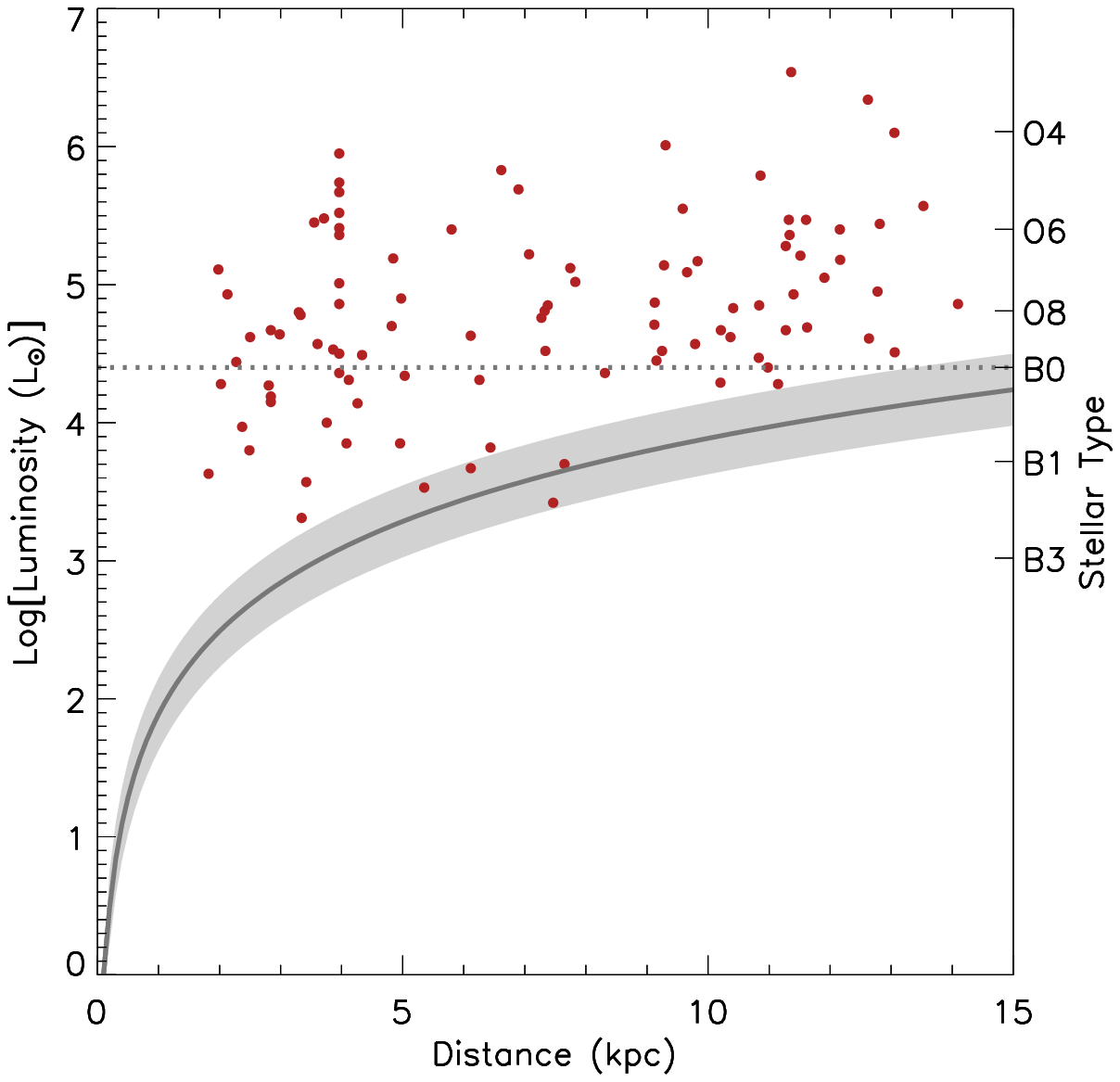}

\caption{\label{fig:luminosities} Luminosity distribution as a function of heliocentric distance. The dark line and light grey shaded region indicates the limiting sensitivity of the MSX 21\,\mum\ band and its associated uncertainty. The sensitivity has been estimated by first calculating the flux in the MSX 21\,\mum\ (band E) using $F_E=4.041\times10^{-14}\,S_{21}$\,Wm$^2$ (\citealt{cohen2000}), where $S_{21}$ is the MSX 21\,\mum\ detection limit ($\sim$2.7\,Jy). This value is then multiplied by a factor of 24, calculated from the ratio of bolometric fluxes, determined from SED fits, to the MSX band E fluxes (\citealt{mottram2011a}). The horizontal dotted line indicates the completeness limit of the RMS survey which corresponds to luminosity of a B0 zero age main sequence star.  }

\end{center}
\end{figure}

In Fig.\,\ref{fig:luminosities} we present a plot showing the distribution of source luminosities as a function of heliocentric distance along with the MSX 21\,\mum\ limiting sensitivity. The number of
sources located at the near and far distances is roughly equal, however, this is a result of the completeness limit since the sensitivity of the 21\,\mum\ MSX band used for our initial selection ($\sim$2.7\,Jy) has resulted in the detection of many more nearby lower-luminosity \HII~regions that fall below the detection threshold at larger distances. The nominal MSX 21\,\mum\ sensitivity corresponds to a detection limit of $\sim$10$^{4.4}$\,\lsun\ at 15\,kpc (\citealt{mottram2011b}). Using this as an estimate of the completeness limit we find the ratio of far to near sources is $\sim$2.3, which is similar to the proportion previously reported (e.g., \citealt{kolpak2003,pandian2008, anderson2009a}).


\section{Summary}
\label{sect:summary}

High-resolution ($\sim$10\arcsec) radio continuum observations were made towards 85 \HII~regions identified by the Red MSX Source (RMS) survey in an effort to resolve kinematic distance ambiguities associated with objects located within the Solar circle. We found the continuum emission was strong enough towards 53 of these \HII~regions to allow \HI\ absorption to be sufficiently detected, and thus, for a distance ambiguity to be resolved. We complement these targeted high-resolution data with 21\,cm spectral line data from the Southern and VLA Galactic plane surveys (SGPS and VGPS respectively) and identified a further 69 \HII~regions with sufficiently strong radio continuum to allow the distance ambiguities to be resolved. In total these two data sets provide continuum information for 105 \HII~regions, with both high and low-resolution data available for 17 sources.

By measuring the velocity of the \HI\ absorption and source velocity
with respect to the velocity of the tangent point we have been able to
resolve the distance ambiguities for 94 \HII~regions and with the aid of additional information drawn from the literature we have assigned distances to a further eight sources. In total we present distances for 102 \HII~regions placing with 39, 60 and 3 sources at the near, far and tangent distances, respectively. Comparing the distances determined for the 17 sources common to both the high and low-resolution data sets we find agreement in only $\sim$65\,per\,cent of cases. This suggests that distances assigned by applying this absorption method to low-resolution data are far less reliable than those assigned using high-resolution data. We also find good agreement with the distances we have determined with those of previous studies reported in the literature.

We investigate the Galactic distribution of this sample of \HII~regions with respect to the large-scale structural features of the Milky Way. Although the sample statistics are too small to expect the \HII~regions to clearly trace spiral-arm structures, we do find the vast majority to be coincident with the expected positions of the far-3\,kpc, Norma and Scutum-Centaurus spiral arms, with the inter-arm regions largely devoid of any \HII~regions. The Galactocentric distribution of southern \HII~regions reveals a single strong peak at approximately 6\,kpc, and is therefore very different to the distribution seen for the RMS northern sample of \HII~regions and MYSOs (\citealt{urquhart2011}). 

In this paper we derive distances and luminosities to a sample of $\sim$102 \HII~regions identified from our programme of follow-up observations designed to examine the global characteristics of this galaxy-wide sample of massive young stars.

\section*{Acknowledgments}

The authors would like to thank the staff of the ATCA for their assistance during the preparation and execution of these observations. We would also like to extend our thanks to the referee for their comments and suggestions, which have helped to improve this manuscript. The Australia Telescope Compact Array is part of the Australia Telescope National Facility which is funded by the Commonwealth of Australia for operation as a National Facility managed by CSIRO. The National Radio Astronomy Observatory is a facility of the National Science Foundation operated under cooperative agreement by Associated Universities, Inc. This research has made use of the SIMBAD database, operated at CDS, Strasbourg, France. This paper made use of information from the Red MSX Source survey database at www.ast.leeds.ac.uk/RMS which was constructed with support from the Science and Technology Facilities Council of the UK.

\bibliography{c1772}

\bibliographystyle{mn2e_jcm}

\end{document}